\documentclass[onecolumn,notitlepage,floatfix,amsmath,amssymb, longbibliography]{revtex4-1}

\usepackage{graphicx}
\usepackage{dcolumn}
\usepackage{color}
\def\ADDA#1{{#1}}       
\def\ADDB#1{{#1}}        
  		
\usepackage{amssymb}
\usepackage[toc,page]{appendix}

\newcommand{\be}{\begin{equation}}
\newcommand{\ee}{\end{equation}}

\usepackage[font=small]{caption}
\usepackage[labelformat = empty,position=top]{subcaption}
\usepackage{graphicx}
\begin{document}
\title{Single-particle dispersion in stably stratified turbulence}

\author{N.E. Sujovolsky$^1$, P.D. Mininni$^1$, and M.P. Rast$^2$}
\affiliation{
      $^1$ Universidad de Buenos Aires, Facultad de Ciencias 
  Exactas y Naturales, Departamento de F\'\i sica, \& IFIBA, CONICET, 
  Ciudad Universitaria, Buenos Aires 1428, Argentina.\\
      $^2$ Department of Astrophysical and Planetary Sciences, 
  Laboratory for Atmospheric and Space Physics, University of 
  Colorado, Boulder, CO 80309,USA}

\begin{abstract}
We present models for single-particle dispersion in vertical and
horizontal directions of stably stratified flows. The model in the
vertical direction is based on the observed Lagrangian spectrum 
of the vertical velocity, while the model in the horizontal direction
is a combination of a continuous-time eddy-constrained random walk
process with a contribution to transport from horizontal
winds. Transport at times larger than the Lagrangian turnover time is
not universal and dependent on these winds. The models yield results
in good agreement with direct numerical simulations of stratified
turbulence, for which single-particle dispersion differs from the well
studied case of homogeneous and isotropic turbulence. 
\end{abstract}
\maketitle

\section{Introduction}

Lagrangian statistics in fluid dynamics offer unique insight into
particle dispersion (e.g., dispersion of pollutants and transport of
nutrients in the ocean) and turbulent mixing 
\cite{yeung_lagrangian_2002, toschi_lagrangian_2009,
  zimmermann_lagrangian_2010, falkovich_lagrangian_2012,
  pumir_single-particle_2016, bec2006effects,
  thalabard2014turbulent}. Often such transport occurs in settings 
such as the atmosphere and oceans, in which turbulence is either
inhomogeneous or non-isotropic due to stratification or 
rotation \cite{wyngaard_atmospheric_1992,biferale_coherent_2016}.
While particle dispersion in homogeneous and isotropic turbulence
(HIT) has received significant attention, particle dispersion in 
anisotropic flows such as stably stratified turbulence has been
studied only recently both in the presence of rotation
\cite{liechtenstein_nonlinear_2005, liechtenstein_role_2006} and
without \cite{kimura_diffusion_1996, aartrijk_single-particle_2008,
  sozza2016large}. Under stable stratification vertical dispersion is
known to be suppressed \cite{kaneda_suppression_2000}, but its effect
on horizontal transport is less certain
\cite{liechtenstein_role_2006,aartrijk_single-particle_2008}.

The study of the role of anisotropies in turbulent mixing is
of central importance in fluid dynamics, as well as in many
geophysical applications. By now it is clear that the presence of
restoring forces such as gravity or rotation, and of their
associated waves, can have a profound impact in the properties of a
turbulent flow which cannot be treated as small corrections to HIT 
\cite{sagaut_cambon_08}. \ADDB{In the particular case of stratified
  turbulence, which plays a key role in geophysics, linear and
  non-linear processes (such as the zig-zag instability
  \cite{billant_self-similarity_2001}, or non-linear resonant
  interactions of internal gravity waves \cite{smith_generation_2002})
  result in a preferential transfer of energy towards vortical modes,
  associated with the development of pancake-like structures (i.e., of
  structures in the flow with typical horizontal scales much larger
  than typical vertical scales), and of strong horizontal winds with
  vertical shear
  \cite{marino_large-scale_2014,clark_di_leoni_absorption_2015}.}
Horizontal turbulent transport in this case can thus be expected to
share similarities with other sheared flows, such as sheared flows
in neutral fluids and in plasmas \cite{terry2000}. Moreover, the
understanding of turbulent transport and mixing in the particular
case of stably stratified turbulence is crucial for atmospheric
sciences and oceanography. In the tropopause, three-dimensional
mixing is believed to play a crucial role in the exchange of
chemical compounds between the stratosphere and the troposphere 
\cite{shapiro1980}. In the oceans, how three-dimensional mixing
develops is important to understand the observed density of
phytoplankton \cite{davis_04}, with implications for the management
of halieutic resources and the fishing industry.

Here we study single-particle statistics in forced stably
stratified turbulence using direct numerical simulations. We show that
for frequencies smaller than the buoyancy frequency, the Lagrangian
vertical velocity follows a spectrum similar to others observed in
wave-dominated flows in the ocean \cite{dasaro_lagrangian_2000}, and
which are often described by an  empirical Garrett-Munk (GM) spectrum
\cite{garrett_space-time_1975, lien_lagrangian_1998}. We then present
models for both vertical and horizontal dispersion. The former
(transport parallel to the mean stratification) indicates that the
reason for the reduced dispersion in this direction is that the flow
is dominated by a random superposition of internal gravity waves. In
the latter (transport perpendicular to the stratification), dispersion
differs from HIT as it is strongly influenced by the large scale
shearing flow generated by the stratification, and which plays an
important role in the atmosphere \cite{smith_generation_2002,
  marino_large-scale_2014, clark_di_leoni_absorption_2015}. The model
used in this case is then a continuous-time eddy-constrained (CTEC)
random walk (which accounts for particle trapping observed in HIT
\cite{rast_turbulent_2016}), with a superposed drift caused by the
vertically sheared horizontal winds (VSHW) in stably stratified
turbulence.

The implications of the models are twofold. On one hand, they provide
a tool to understand fundamental processes affecting turbulent
transport in stratified flows. On the other hand, as the models have
no free parameters and their only ingredients are obtained from
Lagrangian properties of the turbulence or from a knowledge of the
large-scale flow (which in atmospheric and oceanic flows can be
obtained to a good degree by large-scale models), they provide a
statistical way to predict moments of the probability density function
(PDF) of single-particle dispersion without requiring an ensemble of
runs with explicit integration of a large number of tracers. In the
next section we describe the numerical simulations, while in
Sec.~\ref{sec:results} we present the numerical results for vertical
and horizontal dispersion and introduce the models. Finally, we
present our conclusions in Sec.~\ref{sec:conclusions}.

\section{The Boussinesq equations \label{sec:bouss_eq}}

For the numerical simulations we solved the Boussinesq equations for
the velocity ${\bf u}$ and as well as for ``temperature'' fluctuations
$\theta$ (written in units of velocity),
\begin{eqnarray}
\label{eq:n-s_strat}
\frac{\partial {\bf u}}{\partial t} +{\bf u}\cdot{\bf \nabla}{\bf
  u} &=& -{\bf \nabla}p + N \theta {\hat z} +\nu \nabla^{2}{\bf u}+{\bf
  f}, \\
\label{eq:theta}
\frac{\partial \theta}{\partial t}+ {\bf u}\cdot{\bf \nabla} \theta
  &=& -N {\bf u} \cdot {\hat z}  + \kappa \nabla^{2} \theta,
\end{eqnarray}
with the incompressibility condition 
${\bf \nabla}\cdot{\bf u}=0$. Here $p$ is the pressure, $\nu$ the
kinematic viscosity, ${\bf f}$ an external mechanical forcing, $N$
the Brunt-V\"{a}is\"{a}l\"{a} frequency (which sets the background 
stratification), and $\kappa$ the thermal diffusivity. The equations
were solved in a three-dimensional periodic domain of dimensionless
linear length $2\pi$, using a parallel dealiased pseudospectral method
and a second-order Runge-Kutta scheme for time integration
\cite{Mininni11}. All runs have a spatial resolution of $512^3$
regularly spaced grid points, and $\nu=\kappa=8 \times 10^{-4}$ in
dimensionless units (thus, the Schmidt number is 
$\textrm{Sc} = \nu/\kappa = 1$). \ADDB{In all runs described below,
  once the systems reached a turbulent steady state, we injected
  $10^5$ Lagrangian particles distributed randomly in the box. We used
  a high order method to integrate the equations for the Lagrangian
  particles
\begin{equation}
\frac{d {\bf x}_i}{dt} = {\bf u}({\bf x}_i, t) ,
\end{equation}
  (where the subindex $i$ corresponds to the particle label), using a
  second-order Runge-Kutta method in time, and three-dimensional
  third-order spline interpolation to estimate the Lagrangian velocity
  at points that do not correspond to grid points of the fluid code
  (see, e.g., \cite{yeung_algorithm_1988}). Particles move in the
  periodic domain, and thus we assume homogeneity of the turbulent
  flow to re-enter particles that escape out of the domain using
  periodicity.}

The flows were forced at $k=1$ and $2$ with two different
mechanical forcings. A set of two simulations (with different 
Brunt-V\"{a}is\"{a}l\"{a} frequencies, $N=4$ or $N=8$) was forced with
Taylor-Green (TG) forcing \cite{clark_di_leoni_absorption_2015}, which
is a two-component forcing which generates pairs of counter-rotating
von K\'arm\'an swirling flows in planes perpendicular to the
stratification, and with a shear layer in between them. When applied
at only one wavenumber ($k=k_{f}$, the forcing wavenumber), TG forcing
is given by
\begin{equation}
 {\bf f_{TG}}(k_f) = f_{0}\left(sin(k_{f}x)cos(k_{f}y)cos(k_{f}z) ,
 -cos(k_{f}x)sin(k_{f}y)cos(k_{f}z) ,0 \right).
\end{equation}
For $k_f=1$, this forcing has two shear layers, one at $z=\pi/2$ and
another one at $z=3\pi/2$  (where ${\bf f_{TG}}=0$). Our TG forcing,
applied at $k = 1$ and $2$, is simply the superposition 
${\bf f} = {\bf f_{TG}}(1)+{\bf f_{TG}}(2)$. In the presence of
stratification, this mechanical forcing generates a coherent flow at
the large scales, which develops horizontal winds (i.e., a non-zero
mean horizontal velocity) only in the shear layers between the von
K\'arm\'an swirling flows, as the large-scale von K\'arm\'an
structures prevent the formation of strong horizontal winds in the
rest of the domain. Taylor-Green flows, as they excite directly only 
horizontal components of the velocity, have been used before to 
study stratified flows and geophysical turbulence
\cite{riley_dynamics_2003}.

Another set of two simulations (also with Brunt-V\"{a}is\"{a}l\"{a}
frequencies $N=4$ or $8$) was forced using isotropic 
three-dimensional random forcing (RND) with a correlation time of half
a large-scale turnover time. Every $\Delta t = 0.5$, a forcing with
random phases $\phi_{\bf k}$ for each Fourier mode ${\bf k}$ in the
shell $k \in [1,2]$ was generated as
\begin{equation}
{\bf f_1} = f_{0} \sum_{|{\bf k}|\in[1,2]} \Re \left[ i{\bf k}
  \times {\bf \hat{r}} e^{i( {\bf k} \cdot {\bf \hat{r}} + \phi_{\bf
     k} )} \right] ,
\end{equation}
where $\Re$ stands for the real part. The forcing ${\bf f}$ is
obtained by slowly interpolating the forcing from a previous random
state ${\bf f_0}$ to the new random state ${\bf f_1}$, in such a way
that ${\bf f} = {\bf f_1}$ after $\Delta t$. The process is then
repeated to obtain a slowly evolving random forcing which does not
introduce spurious fast time scales in the evolution of the
Lagrangian particles. With this forcing, no large-scale coherent
flows are sustained, and horizontal winds can then grow in the entire
domain. Also, as this forcing is isotropic, a larger amount of
injected power will go into excitation of internal gravity
waves, as confirmed below. Thus, the large scales of both sets of
simulations have very different behaviors. In the Appendix we discuss
a third forcing function, to further validate the model for horizontal
dispersion discussed in the next section using yet another
configuration.

Equations (\ref{eq:n-s_strat}) and (\ref{eq:theta}) have two
control dimensionless parameters. The Reynolds number
\begin{equation}
 \textrm{Re} = \frac{LU}{\nu}, \label{eq:reynolds}
\end{equation}
where $L$ and $U$ are respectively the characteristic Eulerian
length scale and velocity of the flow, and the Froude number
\begin{equation}
 \textrm{Fr} = \frac{U}{LN}, \label{eq:froude}
\end{equation}
which measures the ratio of inertial forces to buoyancy forces in
Eq.~(\ref{eq:n-s_strat}). \ADDB{The characteristic velocity $U$ is
  estimated as the root mean square Eulerian velocity. From the
  Reynolds and Froude numbers, we can also define the buoyancy
  Reynolds number,
  \begin{equation}
    \textrm{Re}_b = \textrm{Re} \, \textrm{Fr}^{2} , \label{eq:reb}
  \end{equation}
  which gives a measure of the strength of the turbulence in the
  stratified flow, and is associated with the turbulent mixing as well
  as with the relevance of viscous effects at all scales in these
  flows (see, e.g., \cite{ivey08}).}

\ADDB{Another useful parameter to estimate the turbulent mixing is
  the local gradient Richardson number (see, e.g.,
  \cite{rosenberg_evidence_2015})
  \begin{equation}
    Ri_{g} = \dfrac{N
      (N-\partial_{z}\theta)}{\partial_{z}(u_{\perp}^{2})} \label{eq:RIG}, 
  \end{equation}
  where $u_{\perp} = (u_{x}^{2}+u_{y}^{2})^{1/2}$. Note that this is a
  pointwise expression. Values of $Ri_{g}<1/4$ can be considered to be
  an indication of possible local shear instabilities at that point,
  while for $Ri_{g}<0$ local overturning can occur.}

In the previous expressions and in the following, the characteristic
Eulerian length scales (or the integral scales) for all runs are
computed from the Eulerian kinetic energy spectrum $E(k)$ of these
flows as
\begin{equation}
  L=2\pi \dfrac{\int{E(k) k^{-1} dk} }{\int{E(k)} dk},
\end{equation}
\begin{equation}
  L_{\parallel}=2\pi \dfrac{\int{E(k_{\parallel}) k_{\parallel}^{-1}
      dk_{\parallel}} }{\int{E(k)} dk},
\end{equation}
and  
\begin{equation}
  L_{\perp}=2\pi \dfrac{\int{E(k_{\perp}) k_{\perp}^{-1} dk} }{\int{E(k)} dk},
\end{equation}
where $L$ is the isotropic Eulerian length scale, $L_{\parallel}$ is
the parallel or vertical Eulerian length scale, and $L_{\perp}$  is
the perpendicular or horizontal Eulerian length scale.

Finally, two other relevant parameters for the next section are the
Eulerian correlation time (or the large-scale turnover time), given by
$T_e=L/U$, and the Lagrangian turnover time $T_{l}$, which is the
mean correlation time of single-particle trajectories. \ADDB{Table 
\ref{tab:char_values} gives the values of the parameters and
characteristic scales introduced in this section for all runs.}

\begin{table}
\centering
\begin{tabular}{p{2.0cm} ccccccccccc}
\hline \hline
Run & Forcing & $N$ & $\nu$ & $\textrm{Re}$ & $\textrm{Fr}$ &
  $\textrm{Re}_{b}$ & $ T_{e}$ & $T_{l}$ & $L$ & $L/L_{\parallel}$ &
  $L/L_{\perp}$ \\
\hline 
TG4 & TG & $4$ & $8\times10^{-4}$ & 7000 & $0.04$ & 11 & $3.6$ &
                                   $7.2$ & $2.2$ & $0.9$ & $0.5$ \\ 
TG8 & TG & $8$ & $8\times10^{-4}$ & 7000 & $0.02$ & 3 & $2.9$ &
                                   $8.0$ & $2.8$ & $1.0$ & $0.5$ \\
RND4 & RND & $4$ & $8\times10^{-4}$ & 13000 & $0.08$ & 83 & $2.0$ &
                                   $25.5$ & $3.5$ & $1.1$ & $0.8$ \\
RND8 & RND & $8$ & $8\times10^{-4}$ & 13000 & $0.04$ & 21 & $2.5$ &
                                   $17.1$ & $3.6$ & $1.1$ & $0.6$ \\
\hline
\end{tabular}
\caption{\ADDB{Parameters and Eulerian and Lagrangian characteristic 
    scales for all runs: TG and RND stand respectively for
    Taylor-Green and random isotropic forcing, $N$ is the
    Brunt-V\"{a}is\"{a}l\"{a} frequency, $\nu$ is the kinematic
    viscosity ($\kappa=\nu$), $\textrm{Re}$ is the  Reynolds number,
    $\textrm{Fr}$ is the Froude number,  $\textrm{Re}_{b}$ is the
    buoyancy Reynolds number, $T_{e}$ and $T_{l}$ are respectively the
    characteristic Eulerian and Lagrangian time scales, $L$ is the
    isotropic Eulerian integral length scale, and $L/L_\parallel$ and
    $L/L_{\perp}$ are the ratios of this scale respectively to the
    parallel and to the perpendicular integral length scales.}}
\label{tab:char_values}
\end{table}
\section{Results \label{sec:results}} 

\subsection{Particle trajectories}

\begin{figure}
\centering
\includegraphics[width=7.3cm]{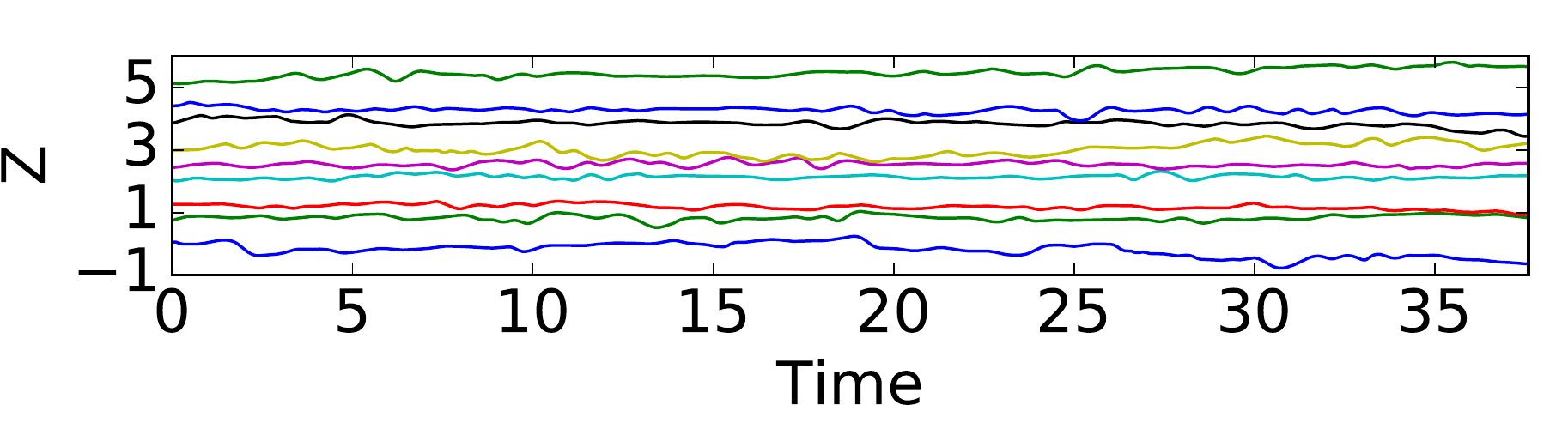}   
\hskip .3cm
\includegraphics[width=7.3cm]{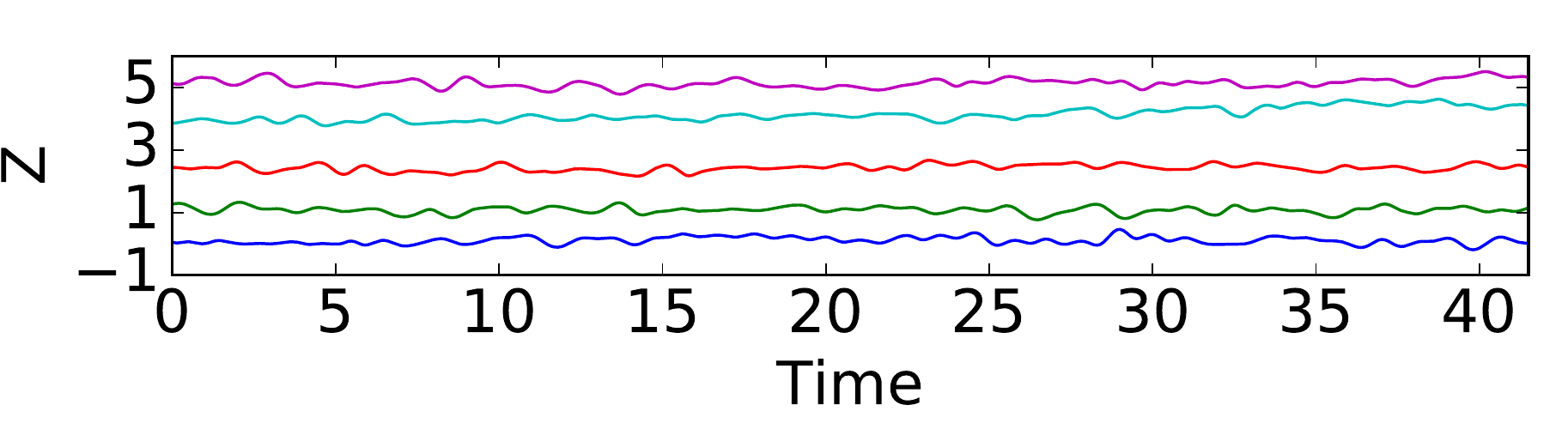}   
\includegraphics[width=7.6 cm]{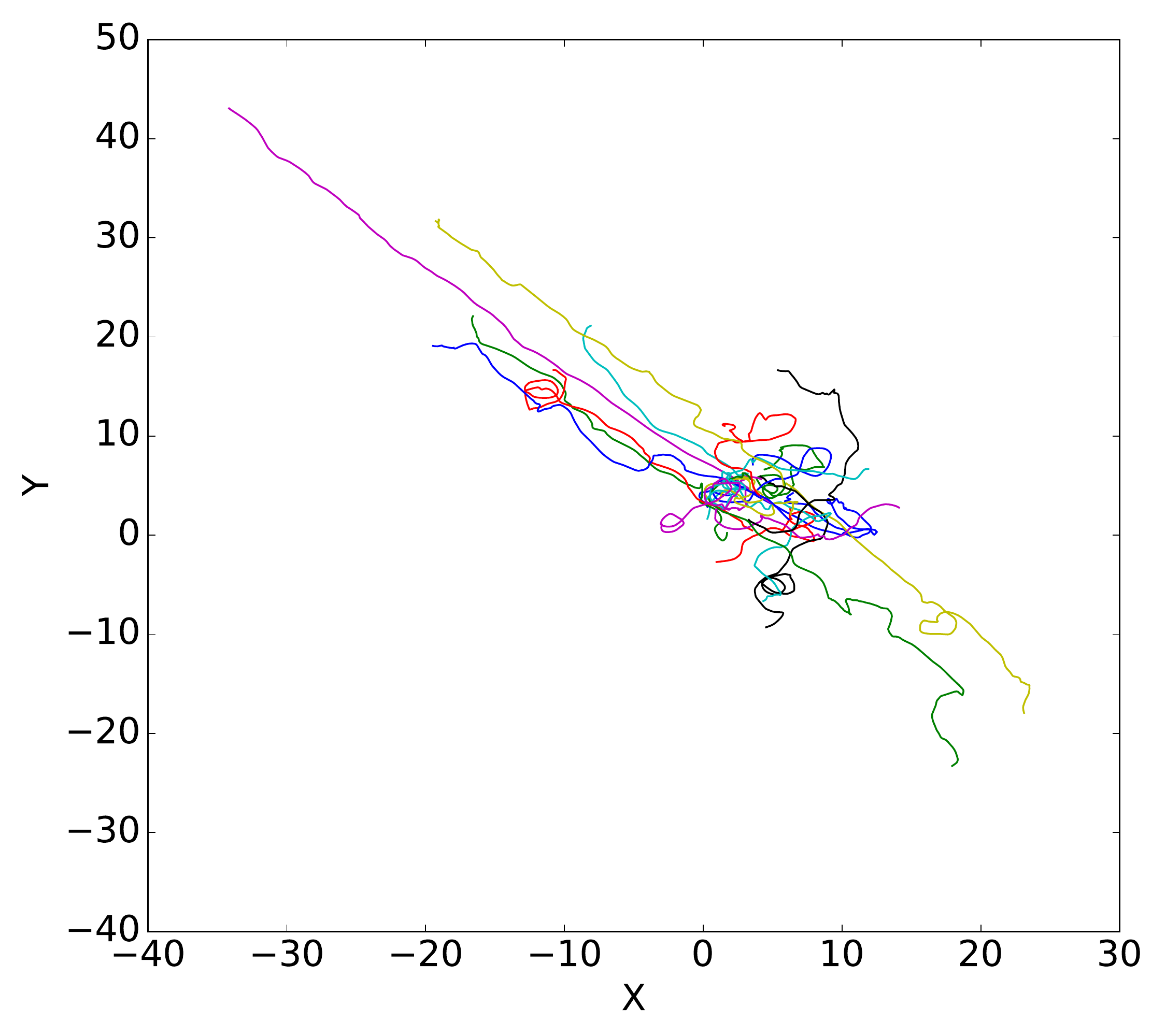}  
\includegraphics[width=7.6 cm]{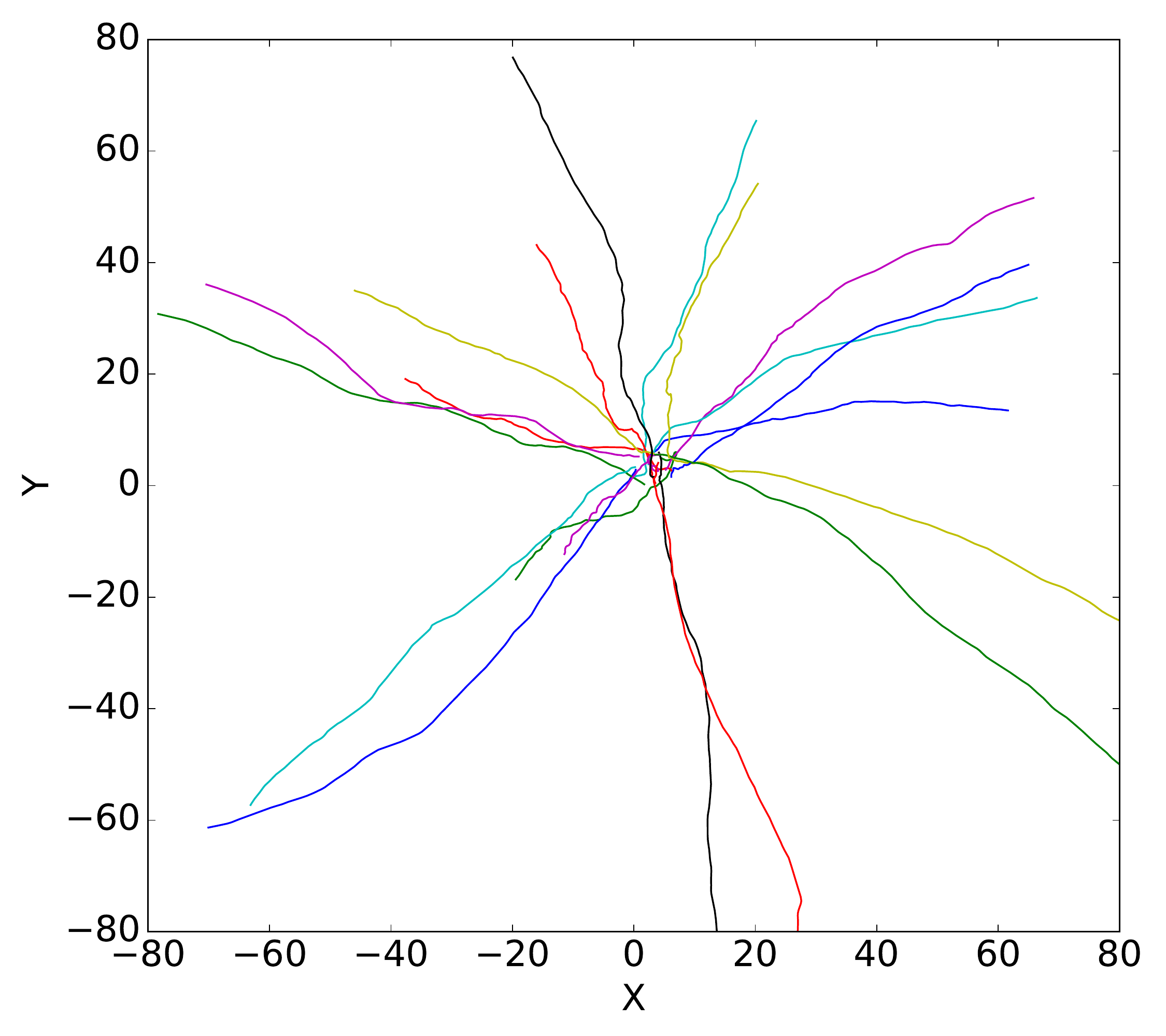}  
\caption{({\it Color online}) {\it Left column:} Vertical ({\it top})
  and horizontal ({\it bottom}) displacements for a few particles
  (indicated by different colors) in the TG4 run (Taylor-Green forcing
  with $N=4$). {\it Right column}: Same for the RND4 run (random
  forcing with $N=4$).}
\label{fig:xyz}
\end{figure}

Figure \ref{fig:xyz} shows vertical and horizontal displacements for a
few particles in the RND4 simulation (random forcing with $N=4$;
in the following we label runs by their forcing followed by the value of
$N$, following the notation in table \ref{tab:char_values}), and in
the TG4 simulation (i.e., Taylor-Green forcing with $N=4$). In both
simulations, while vertical displacements are small and display
wave-like motions, horizontal motions are large (compared with the
periodic domain size of $2\pi$). In the RND4 simulation horizontal
trajectories seem almost ballistic at all times, being dominated by a
strong drift. Note also that particles at different heights (indicated
by the different colors) move in different directions, as the
horizontal winds in each layer also point in different directions. In
the TG4 simulation, horizontal displacements show only a fraction of
the particles with such a drift (those in the vicinity of the
horizontal layers where Taylor-Green forcing is zero, and all moving
along the same direction), and a significantly stronger trapping of
particles by eddies which can be seen as particle trajectories turn
around.

\subsection{Lagrangian spectra}

\begin{figure}
\centering
\includegraphics[width=8.5cm]{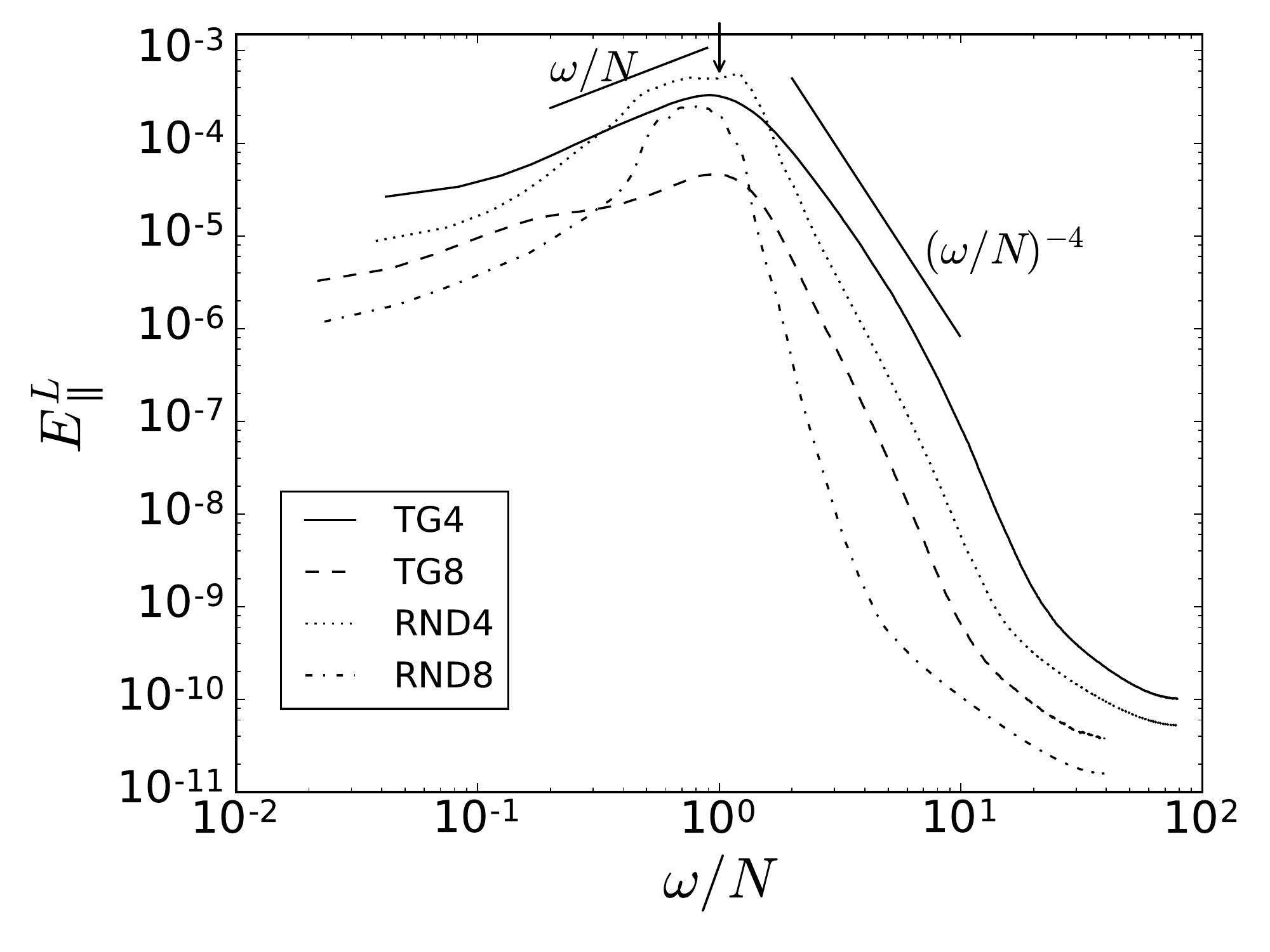}
\includegraphics[width=8.5cm]{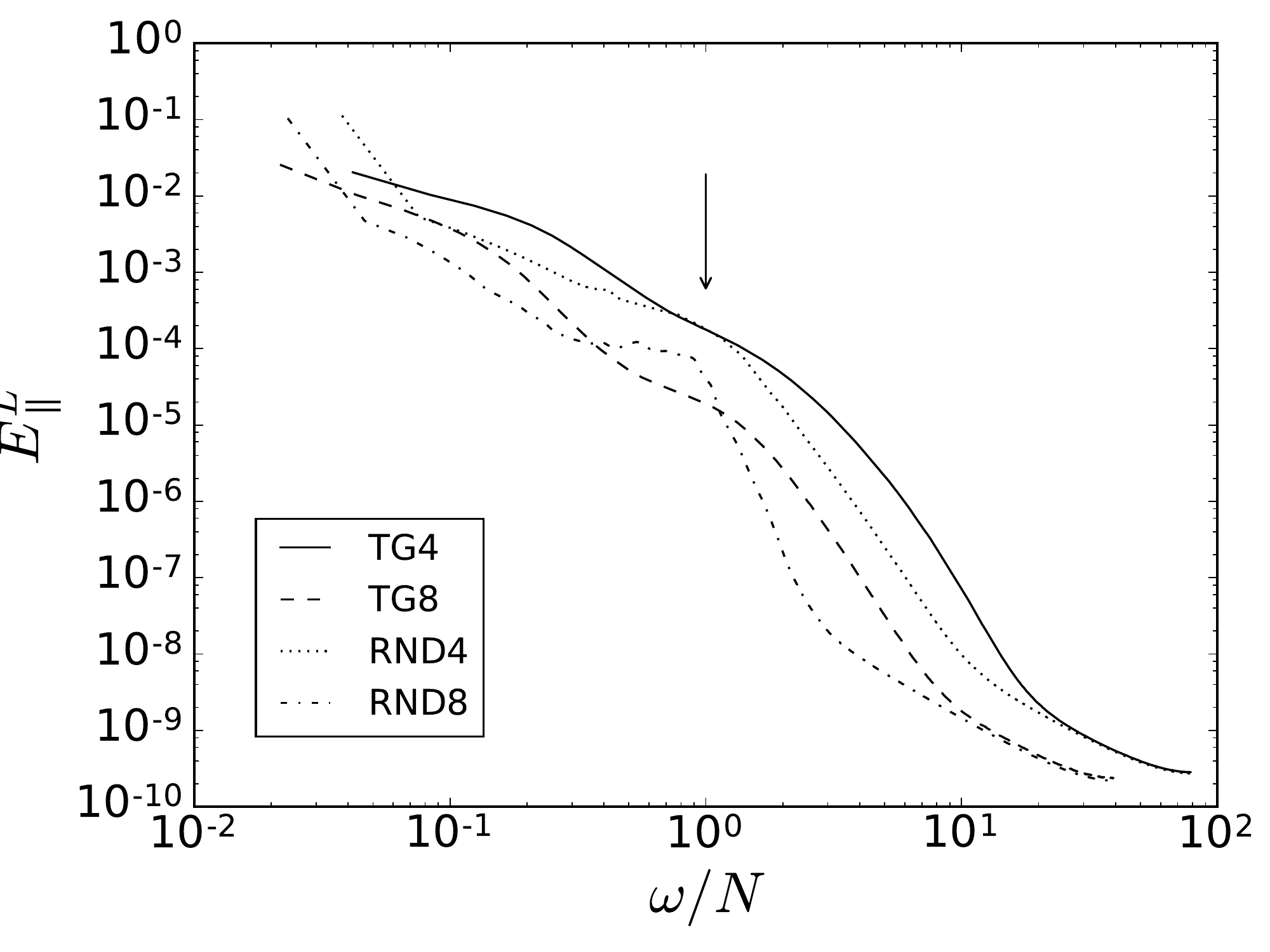}
\caption{({\it Color online}) {\it Left}: Lagrangian power spectrum of
  the vertical velocity in all simulations. Power laws are indicated
  as a reference (see Table \ref{tab:exponents}). All spectra display
  a peak at the Brunt-V\"{a}is\"{a}l\"{a} frequency (indicated by the
  arrow), and for frequencies $\omega/N<1$ the shallow spectra are 
  reminiscent of observations of oceanic internal gravity waves. 
  {\it Right}: Lagrangian power spectrum of the horizontal  
  velocity. In the horizontal case there is no clear peak near the
  Brunt-V\"{a}is\"{a}l\"{a} frequency.}
\label{fig:vspec_all}
\end{figure}

The Lagrangian spectrum of the vertical velocity is shown in
Fig.~\ref{fig:vspec_all}. The spectra of all simulations are shallow
for frequencies $\omega/N < 1$, and show a peak near $\omega/N=1$,
i.e., the parallel kinetic energy is concentrated near the buoyancy
frequency. \ADDB{The shallow spectra for frequencies $\omega/N < 1$
  are reminiscent of the GM spectrum, which is an empirical spectrum
  for internal oceanic waves \cite{garrett_space-time_1975,
    lien_lagrangian_1998, dasaro_lagrangian_2000} that reflects the
  dominance of wave contributions. The spectrum was first derived for
  the total energy, but later generalized for vertical velocities
  \cite{garrett_space-time_1975}, and used to compare with Lagrangian
  measurements (see, e.g., \cite{dasaro_lagrangian_2000,
    d2007high}). In the absence of rotation, i.e., at oceanic scales
  much smaller than the scales at which rotation is relevant, the
  spectrum reduces to a white spectrum for $\omega \le N$.} For
frequencies $\omega/N > 1$ the vertical Lagrangian spectra in
Fig.~\ref{fig:vspec_all} decrease rapidly (in some cases following a
power law). Only as a reference, in Fig.~\ref{fig:vspec_all} 
({\it left}) we show two power laws, for frequencies 
$\omega/N \lesssim 1$, and for $\omega/N > 1$. 

\ADDB{On this light, we can interpret the observed vertical Lagrangian
  spectrum as follows. For frequencies  $\omega/N < 1$, the shallow
  spectrum corresponds to a superposition of internal gravity waves
  and is dominated by their contributions. Indeed, as the dispersion
  relation of internal gravity waves is  $\omega = N k_\perp/k \le N$,
  waves can only contribute to frequencies with $\omega/N \le 1$. For
  $\omega/N > 1$ waves cannot contribute, and any power at those
  frequencies must come from fast vortical motions and turbulence, as
  also observed in oceanic measurements where power laws were also
  reported for frequencies larger than the buoyancy frequency
  \cite{dasaro_lagrangian_2000}. Moreover, note that the RND 
  simulations have steeper spectra at frequencies larger than the
  Brunt-V\"{a}is\"{a}l\"{a} frequency, even though they have larger
  $\textrm{Fr}$ and $\textrm{Re}_b$ numbers. This is compatible with a
  stronger predominance of waves in those runs, associated with the
  direct excitation of waves by the isotropic three-dimensional
  forcing. The stronger waves and the weaker turbulence in these runs
  will be confirmed later when we study the gradient Richardson number
  and we show that the TG simulations have larger probabilities of
  satisfying the conditions for local shear instabilities or
  overturning.}

In all cases, the predominance of waves in the vertical Lagrangian
spectra, which concentrate most of the  power, is compatible with the
vertical Lagrangian trajectories observed in Fig. \ref{fig:xyz}: there
is little dispersion in this direction, and particle displacements are
dominated by wave-like motions. In comparison, Lagrangian spectra of
the horizontal velocity (see Fig.~\ref{fig:vspec_all}) do not display
a peak at the buoyancy frequency nor a shallow spectrum for
frequencies $\omega/N < 1$, although in some simulations the
horizontal spectra display a knee and a change in the spectral slope
in the vicinity of this frequency (especially, again, for RND
forcing).

\begin{figure}
\centering
\includegraphics[width=8.5cm]{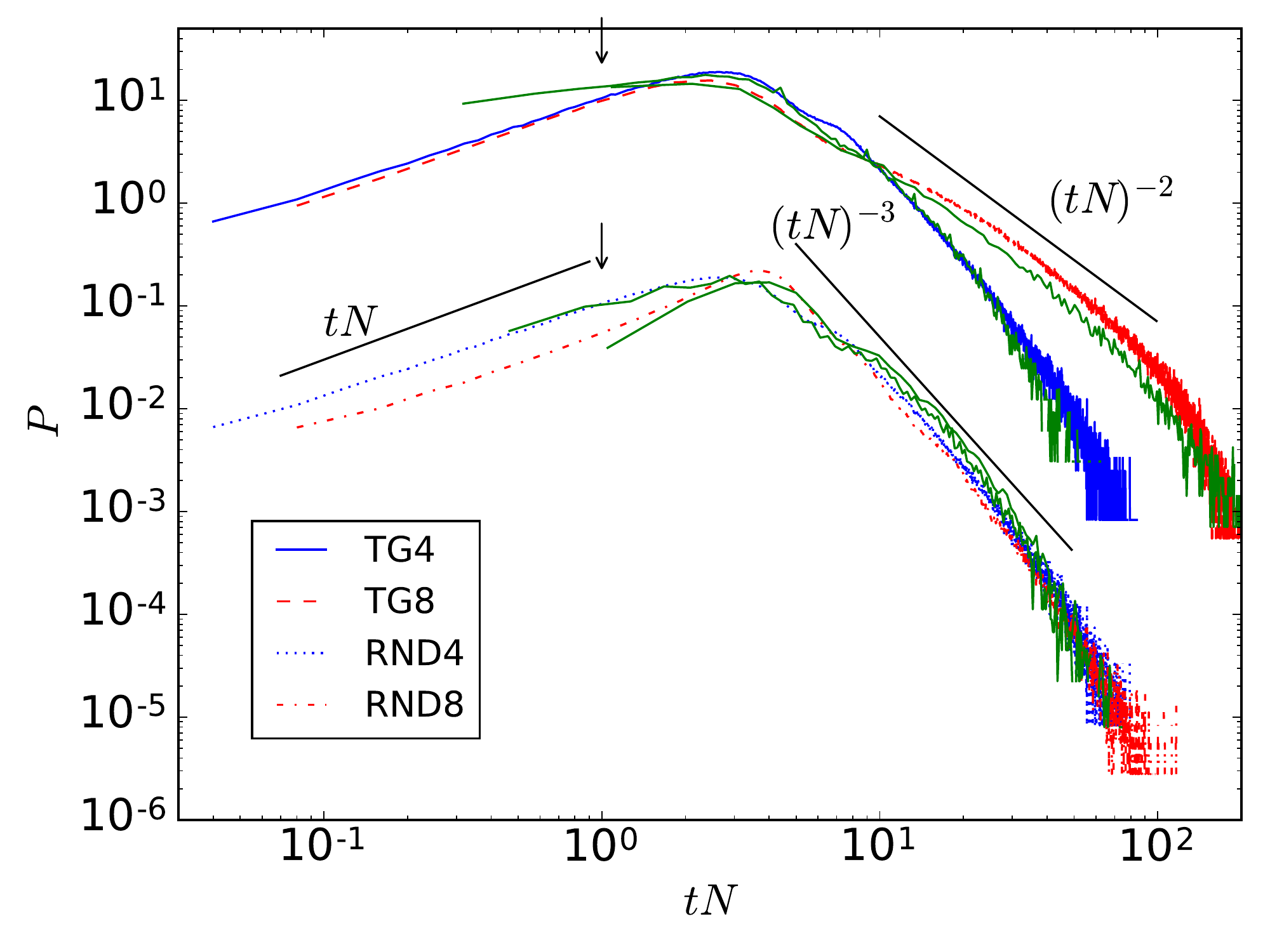}
\caption{({\it Color online}) PDF of vertical displacement waiting
  times for the simulations, and waiting times generated by the model
  consisting of a superposition of random waves (solid green in all
  cases). For TG runs the PDFs have been shifted vertically for better
  visualization. Power laws are shown as a reference (see Table
  \ref{tab:exponents}), and vertical arrows indicate $t = 1/N$.} 
\label{fig:WT}
\end{figure}

\subsection{Vertical dispersion}

\begin{figure}
\centering
\includegraphics[width=8.5cm]{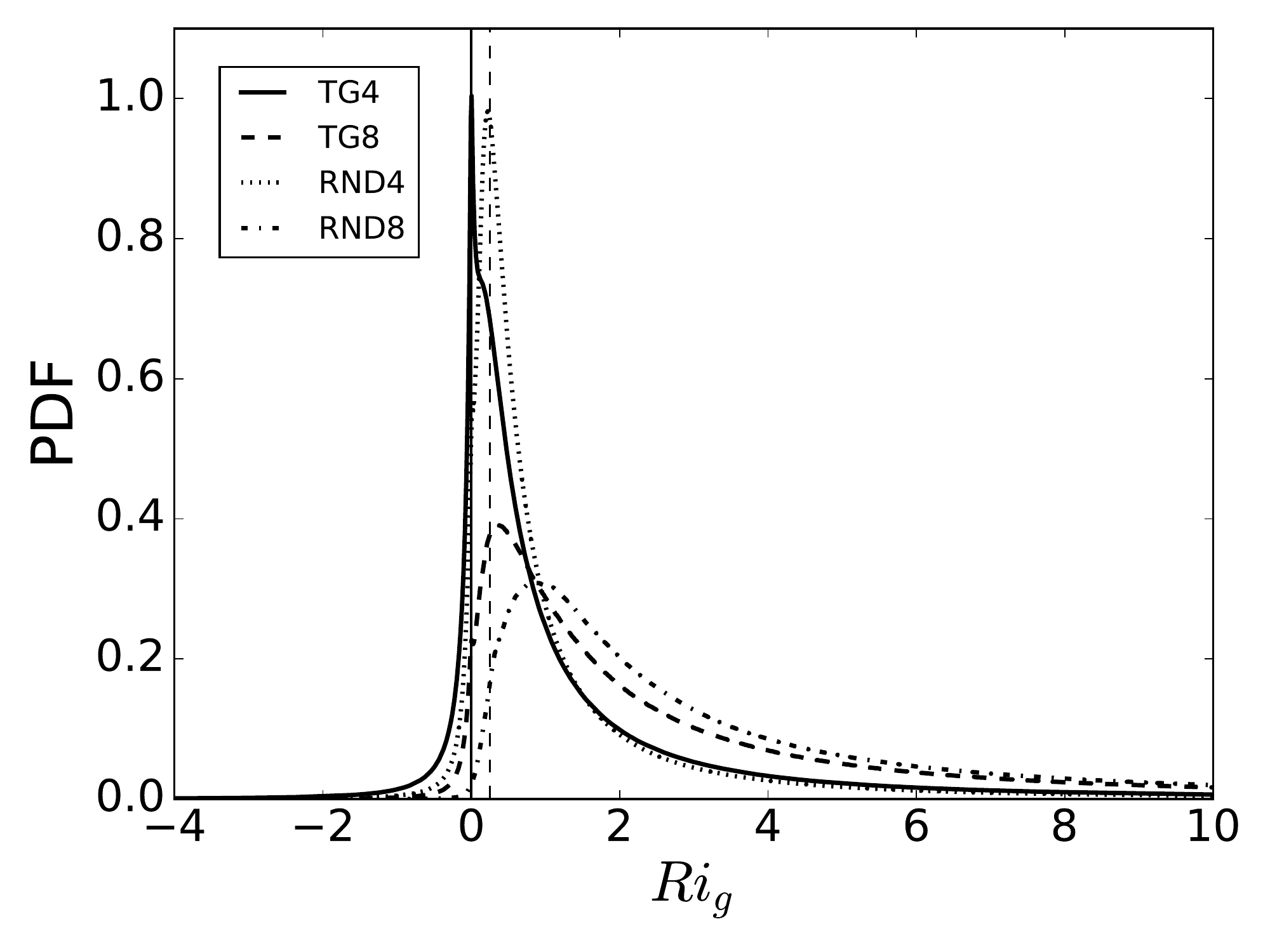}
\caption{\ADDB{PDFs of the local gradient Richardson number $Ri_g$ for
    all runs. A vertical solid line at $Ri_{g}=0$ and a vertical
    dashed line at $Ri_{g}=1/4$ are shown as references; for
    $Ri_{g}<1/4$ local shear instabilities can develop, while for
    $Ri_{g}<0$ overturning can occur.}}
\label{fig:RIG}
\end{figure}

To model the small displacements in the vertical direction, we
consider the statistics of vertical displacement waiting times of
the Lagrangian particles. To this end, we take the waiting time as the
time interval between two consecutive crossings of each particle
trajectory through its mean $z$ elevation. We plot the waiting time
distribution for all runs in Fig.~\ref{fig:WT}. The PDFs are
non-exponential, indicating the system has memory (waves carry
information of their initial conditions for a finite amount of time),
and are compatible with a power law for $tN \gtrsim 1$. The slope is
steeper for the RND runs, $\sim t^{-3}$, than for those with TG
forcing,  $\sim t^{-2}$ (the best fit to these exponents, $\beta_-$,
corresponding to the exponents for $tN>1$, as well as for exponents
$\beta_+$ corresponding to power laws approximated for $tN<1$, are
shown in table \ref{tab:exponents}). Surprisingly, in the TG case, the
probability of long waiting times (i.e., of large excursions of some
particles from their mean $z$ position) also increases with increasing
stratification. This reflects the underlying nature of the stratified
turbulence, in which waves can be nonlinearly amplified 
\cite{rorai_turbulence_2014} (note long excursions can be generated by
strong low frequency waves, as observed in Fig.~\ref{fig:vspec_all}
and in the model shown next), which can also result in a local
instability or in overturning. \ADDB{To confirm this,
  Fig.~\ref{fig:RIG} shows the PDFs of the local gradient Richardson
  number $Ri_{g}$ for all runs. When $Ri_{g}<1/4$ local shear
  instabilities can develop, while for $Ri_{g}<0$ overturning is
  possible as the vertical gradient of temperature fluctuations
  $\partial_z \theta$ overcomes the background gradient associated
  with $N$. The TG runs have larger probability of having points with
  $Ri_{g}<1/4$ or $Ri_{g}<0$ when compared with the RND runs at fixed
  Brunt-V\"{a}is\"{a}l\"{a} frequency. Also, as stratification (or
  equivalently, $N$) is increased, the PDFs are shifted to the right,
  resulting in lower probability of local shear instabilities or
  overturning. However, note this effect is stronger in the RND runs
  when compared with the TG runs. In particular, for the TG8 run there
  is still a considerable $(2.8 \pm 0.1)\%$ probability of finding
  points with $Ri_{g} < 0$, while for the RND8 run the probability is
  $(0.2 \pm 0.1) \%$. This is compatible with the observations made
  before about the stronger prevalence of waves in the RND runs (which
  have larger values of $\textrm{Re}_b$, but are forced isotropically)
  when studying the Lagrangian spectrum of the vertical velocity.}

Based on these observations, we thus propose a simple model for the
vertical particle motion, using a superposition of waves with
random phases, and with amplitudes determined by the observed parallel
Lagrangian velocity  $v_{\parallel}^{L}$ or by its Lagrangian spectrum
$E_{\parallel}^{L}(\omega)$ on dimensional grounds as
\begin{equation}
A(\omega) \sim v_{\parallel}^{L}(\omega)/\omega \sim
    \left. \sqrt{\omega \, E_{\parallel}^{L}(\omega)} \right/\omega 
    \sim \omega ^{(\alpha_\pm-1)/2} .
\label{eq:amplitude}
\end{equation}
Note $E_{\parallel}^{L}(\omega)$ is approximated by two power laws:
one for $\omega \lesssim N$ (with exponent $\alpha_+>0$, see 
Fig.~\ref{fig:vspec_all}), and one for $\omega > N$ (with exponent
$\alpha_-<0$). The values of the exponents $\alpha_\pm$ obtained from
a best fit to the Lagrangian vertical spectra in
Fig.~\ref{fig:vspec_all} are shown in Table \ref{tab:exponents}.  The
waiting times generated by this random superposition of waves are also
shown in Fig.~\ref{fig:WT}. The good agreement between the model and
the data indicates that, at least for the Froude and Reynolds numbers
considered here, the quenching of vertical dispersion results from the
dominance of waves, and that the empirical knowledge of the turbulent
vertical Lagrangian spectrum is sufficient to predict the probability
of vertical excursions by the Lagrangian particles (at least for times
comparable to several periods of the waves).

\begin{table}
\centering
\setlength{\tabcolsep}{1.0em}
\begin{tabular}{p{2.0cm} c c c c}
\hline \hline
Run & $\alpha_{+}$ & $\alpha_{-}$ & $\beta_{+}$ & $ \beta_{-}$ \\
\hline
TG4  & $0.8 \pm 0.3 $ &  $-4.1 \pm 0.1$ &  $0.8\pm0.1$ & $-2.9\pm0.1$ \\
TG8  & $0.8 \pm 0.2 $ &  $-5.5 \pm 0.1$ &  $0.9\pm0.1$ & $-2.0\pm0.1$ \\
RND4& $1.3 \pm 0.4 $ &  $-5.3 \pm 0.1$ &  $0.9\pm0.1$ & $-3.3\pm0.1$ \\
RND8& $1.0 \pm 0.2 $ &  $-9.0 \pm 0.1$ &  $1.0\pm0.1$ & $-3.3\pm0.1$ \\
\hline
\end{tabular}
\caption{Exponents of power laws in the Lagrangian parallel
    spectrum and in the waiting time distributions for all runs, with
    error bars. $\alpha_{\pm}$ are the exponents of power laws
    obtained from a best fit to the Lagrangian parallel spectra in
    Fig.~\ref{fig:vspec_all}, with $\alpha_+$ corresponding to the
    exponents for $\omega/N\lesssim 1$, and $\alpha_-$ to the
    exponents approximated for $\omega/N>1$. $\beta_{\pm}$ are the
    exponents of power laws approximated for the waiting time
    distributions in Fig.~\ref{fig:WT}, with $\beta_+$ corresponding
    to the exponents for $tN<1$, and $\beta_-$ to the exponents for
    $tN>1$.}
\label{tab:exponents}
\end{table}

\begin{figure}
\centering
\includegraphics[width=8.5cm]{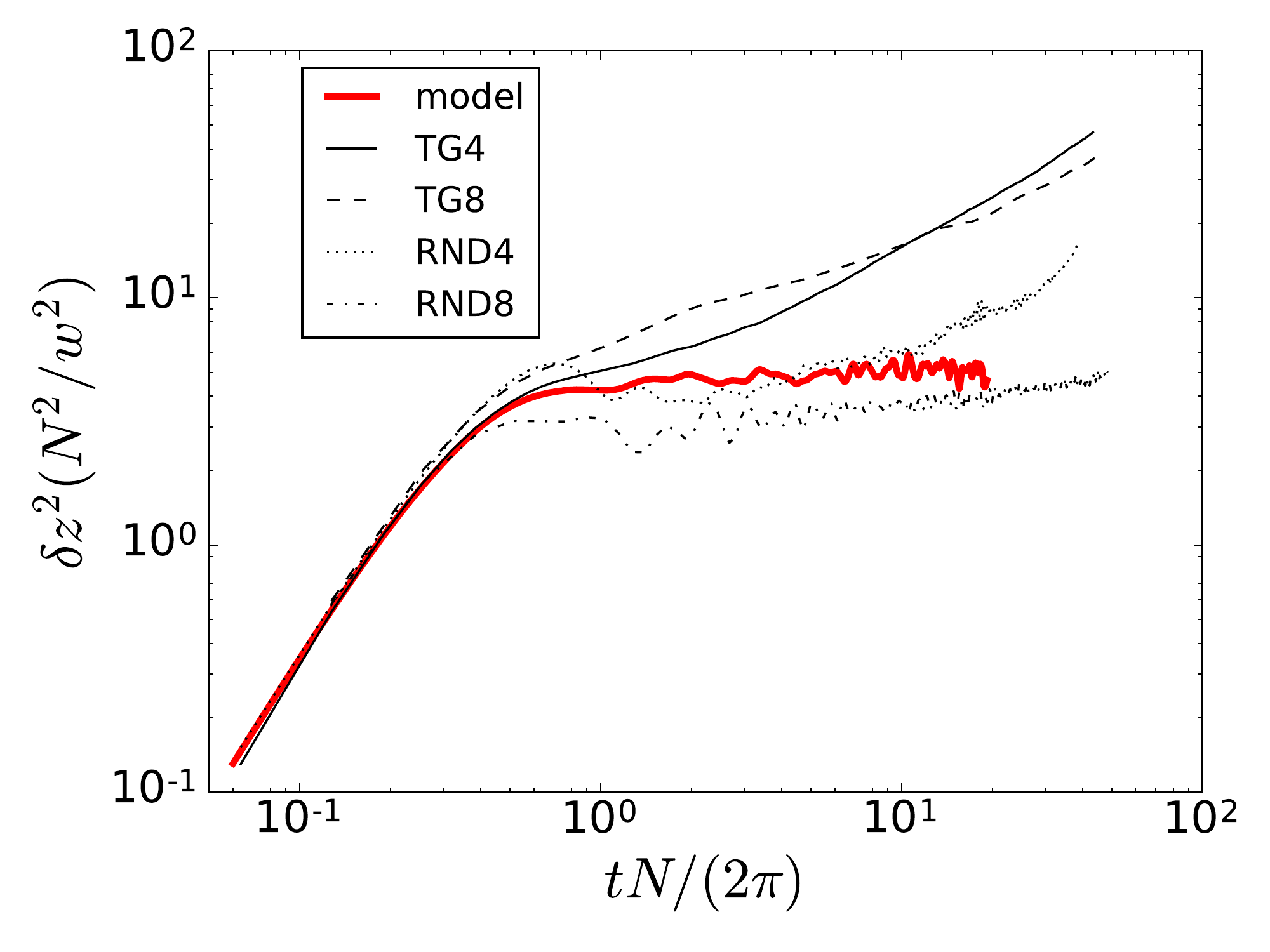}
\caption{({\it Color online}) Mean vertical quadratic dispersion
    $\delta z^2$ for all runs, normalized by the ratio $w^2/N^2$ of
    the mean squared vertical Lagrangian velocity to the squared
    Brunt-V\"{a}is\"{a}l\"{a} frequency. Time is normalized by the
    Brunt-V\"{a}is\"{a}l\"{a} period. All simulations collapse at
    early times, but the TG runs depart after $tN/2\pi \approx 1$. 
    The thick (red) curve shows the model for RND4, which is in 
    good agreement with the simulation until at late times molecular
    diffusion results in a slow vertical dispersion of the particles.}
\label{fig:vertical_disp}
\end{figure}

In \cite{aartrijk_single-particle_2008}, a normalization of the
mean vertical quadratic displacement of the particles $\delta z^2$
was proposed to reobtain (at least for early times) a behavior
similar to that found in HIT. In Fig.~\ref{fig:vertical_disp} we
show $\delta z^2$ normalized in this way, namely by multiplying the
vertical quadratic dispersion by the ratio $N^2/w^2$ of the squared
Brunt-V\"{a}is\"{a}l\"{a} frequency to the mean squared vertical
Lagrangian velocity (with 
$w^2 = \langle (v_{\parallel}^{L})^2 \rangle$), and by multiplying
time by $N/2\pi$. The quadratic vertical dispersions of all runs
collapse to a single curve for $tN/2\pi \lesssim 1$, confirming the
scaling proposed in \cite{aartrijk_single-particle_2008} up to the
period of the slowest internal gravity waves, and for longer times
for RND forcing. In \cite{nicolleau_turbulent_2000}, a kinematic
model also based on a random superposition of waves was presented to
explain the observed vertical quadratic displacement, which also
results in a fast growth up to the buoyancy period, and in saturation
for later times. As discussed in
\cite{aartrijk_single-particle_2008} (which studied simulations 
with random forcing), the slow dispersion at even later times in the
RND runs is probably due to molecular diffusion. In
Fig.~\ref{fig:vertical_disp} we also show $\delta z^2$ constructed
from our model using the parameters of run RND4, which is also in 
good agreement with the data. The case of TG forcing is different
from the RND runs and from the behavior reported in
\cite{nicolleau_turbulent_2000, aartrijk_single-particle_2008}, as
vertical dispersion continues to grow with time after $tN/2\pi =
1$. As discussed above, these simulations display larger probabilities
of long waiting times (i.e., of long vertical excursions of the
particles), which are associated with the development of local
overturning in the flow. \ADDB{This is indeed what the PDFs of
  $Ri_{g}$ show (see Fig. \ref{fig:RIG}), and suggests that
  modifications to this model would be required if the turbulence is
  increased further, or if the stratification is decreased. A
  parametric study of vertical dispersion varying $\textrm{Re}_b$ for
  this forcing is left for a future study.} Finally, note that
although at early times in all runs the scaling $\delta z^2 \sim t^2$
could suggest that for short times the system behaves like HIT, the
agreement with our model (as well as the validity of this scaling only
up to the Brunt-V\"{a}is\"{a}l\"{a} period) indicates that this
behavior, at least for the range of parameters considered in this
study, is caused by the vertical transport of particles by the random
superposition of internal gravity waves.

\subsection{Horizontal dispersion}

Dispersion of Lagrangian particles in the horizontal direction can be
large (see Fig.~\ref{fig:xyz}), and at first sight (at least for the
TG runs) it can appear to be similar to that of HIT (see also
\cite{aartrijk_single-particle_2008}). Since vertical dispersion is
small, particle motions in planes perpendicular to the mean
stratification can be approximated as two dimensional, and 
prediction of dispersion in this direction is relevant for the
stably stratified atmosphere and for other geophysical flows. Thus, a
model for the probability distribution $P({\bf x},t;{\bf x'},t')$ of
finding a particle at $({\bf x},t)$ given a previous location 
$({\bf x'},t')$ has multiple applications, and would allow
probabilistic prediction of the concentration of quantities
transported by the flow without resorting to ensembles of
deterministic simulations with small differences in the initial 
concentrations \cite{rast_turbulent_2016}. In the following we derive
a model for this distribution resorting only to general statistical
properties of the turbulent flow.

\begin{figure}
\hskip .2cm
\includegraphics[width=8.5cm]{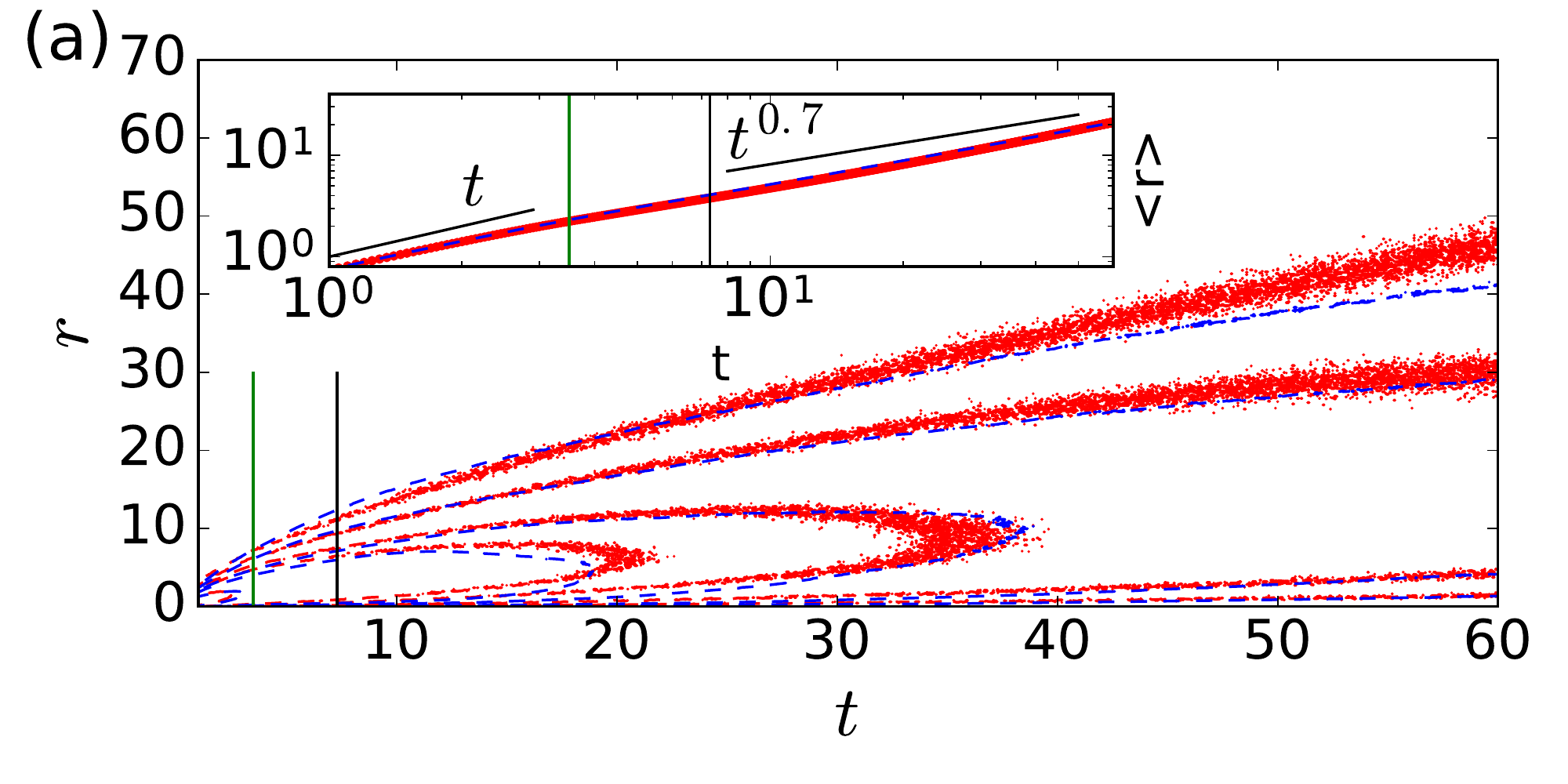}
\includegraphics[width=8.5cm]{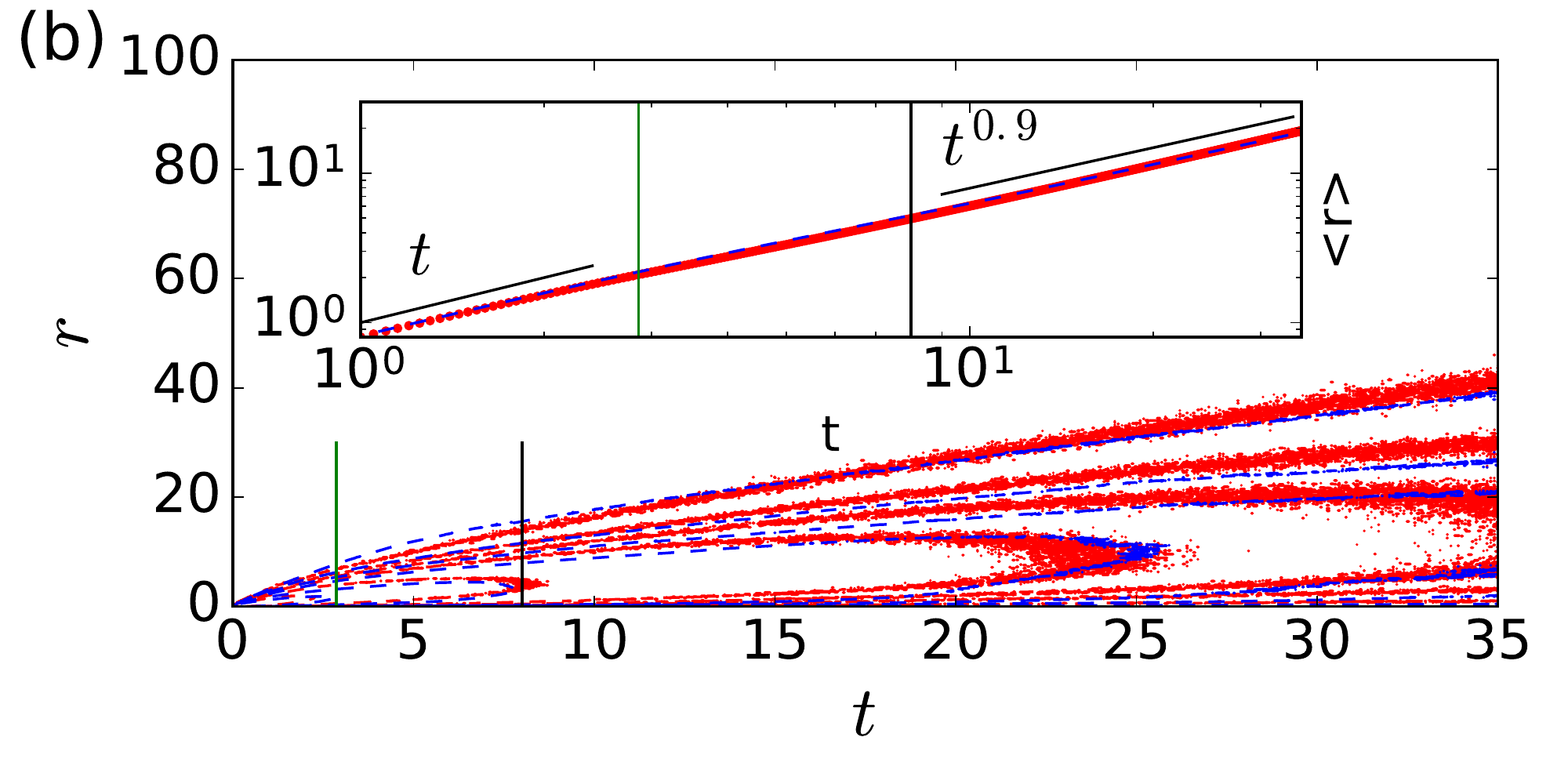}
\includegraphics[width=8.5cm]{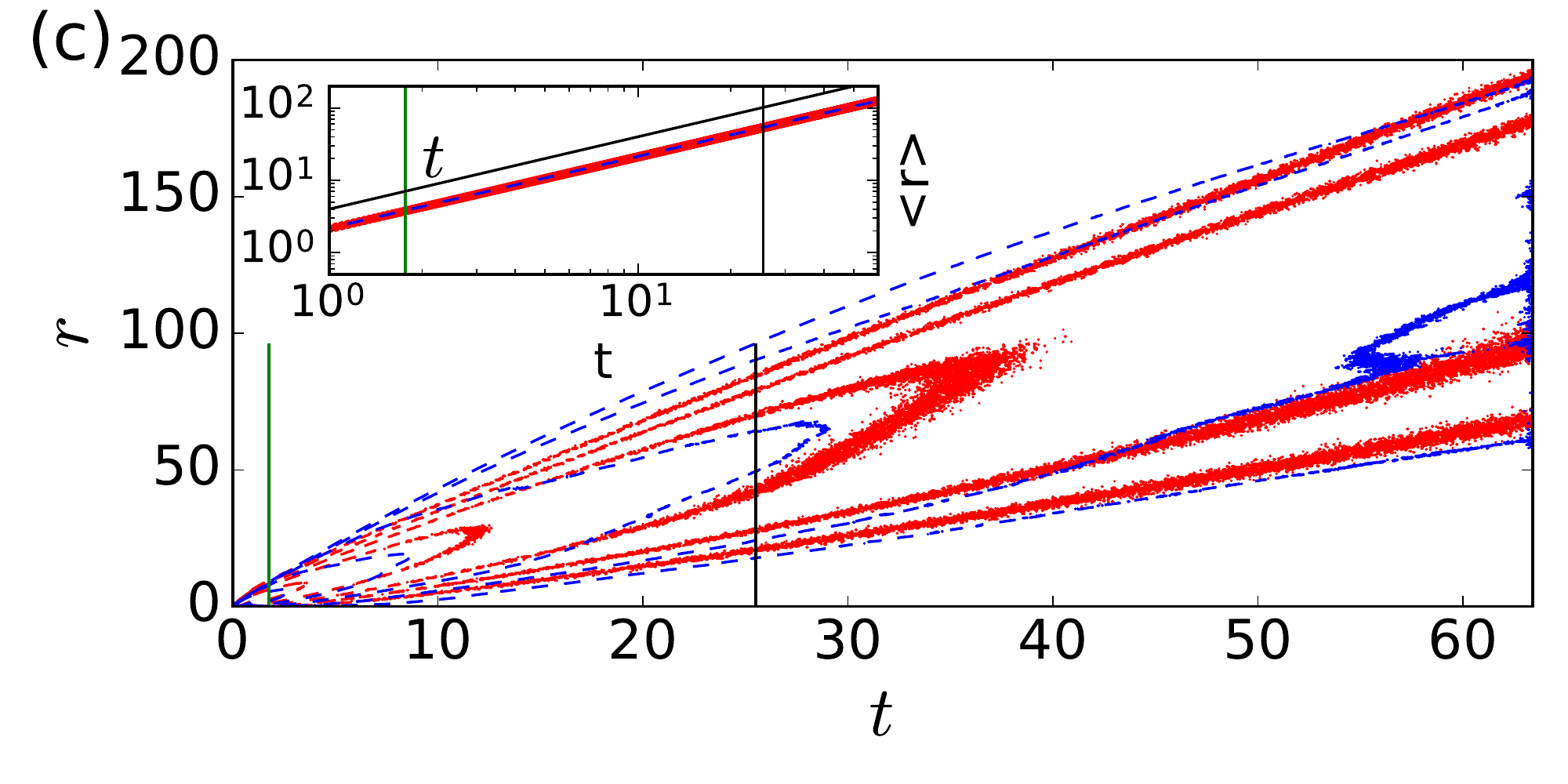}
\includegraphics[width=8.5cm]{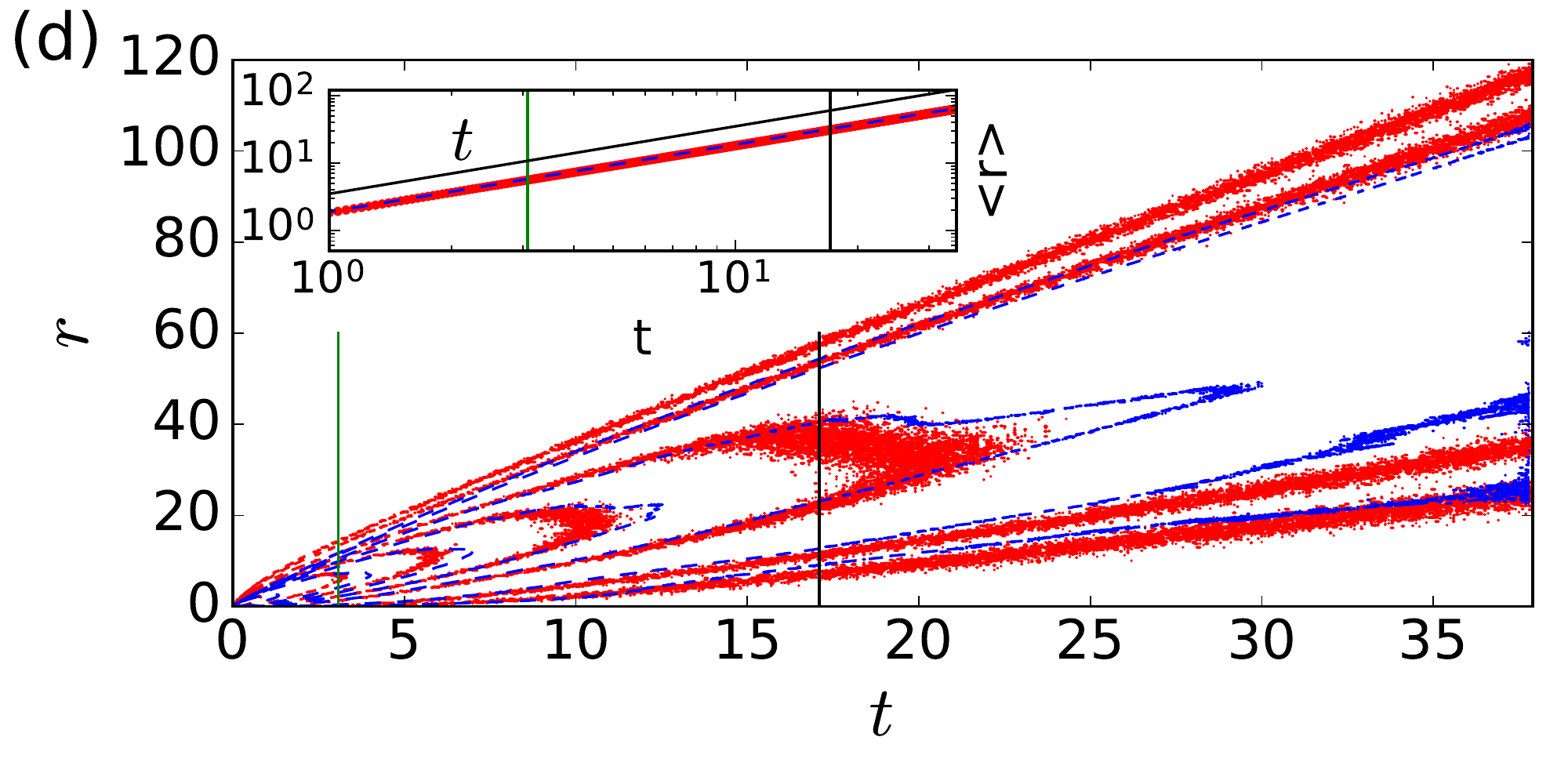}
\caption{{\it (Color online)} Isocontours of $P(r,t)$ for the
  simulations (dashed blue lines) and for the model (dotted red
  lines), for (a) TG4, (b) TG8, (c) RND4, and (d) RND8. The inset show
  the mean horizontal displacement with the same line labels (power
  laws are shown as references, see text for details). Vertical 
  lines (from left to right in all figures) indicate respectively the
  Eulerian $T_{e}$ and Lagrangian $T_{l}$ turnover times.}
\label{fig:prt} 
\end{figure}

\begin{figure}
\centering
\includegraphics[width=8.5cm]{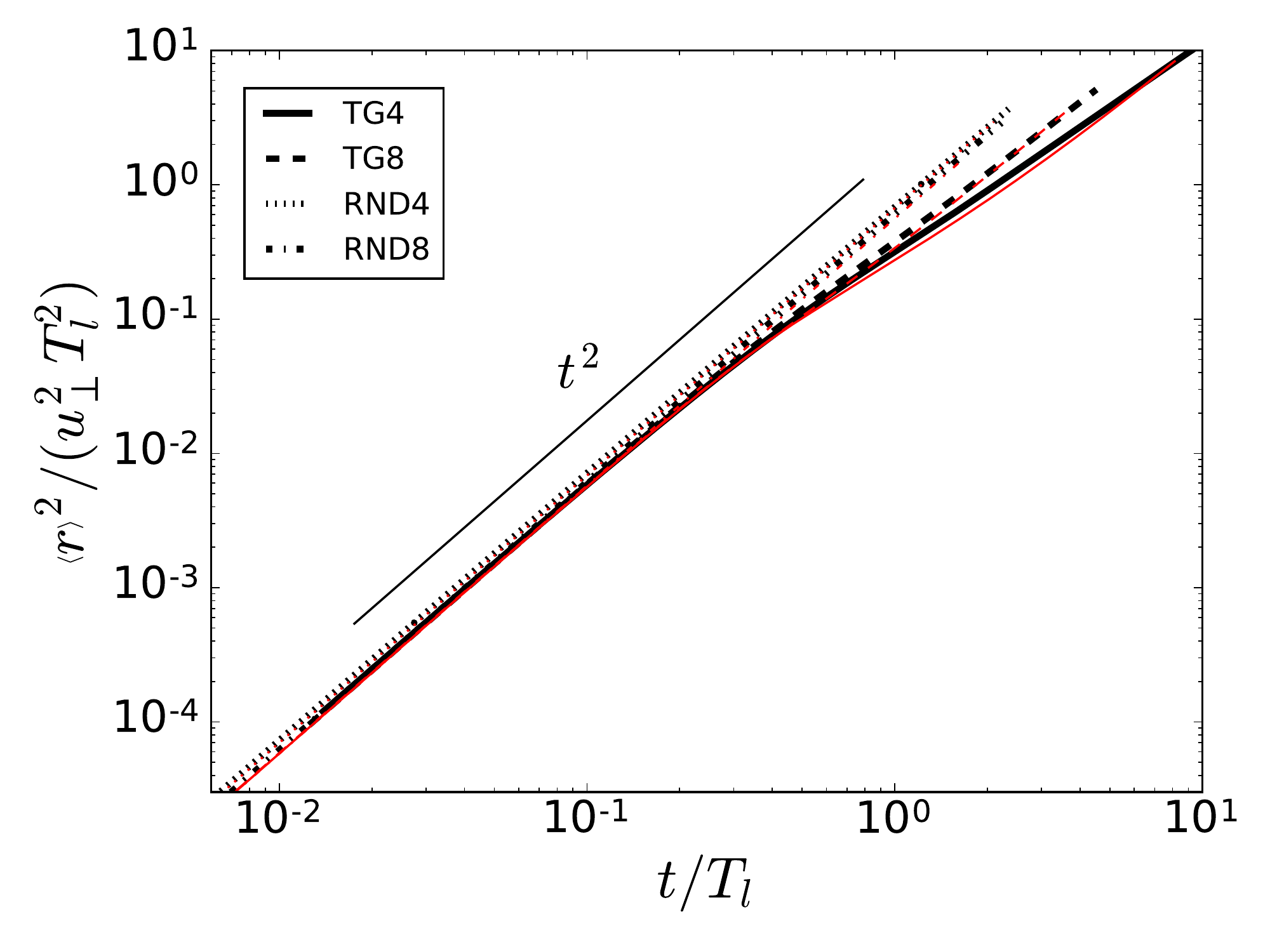}
\includegraphics[width=8.5cm]{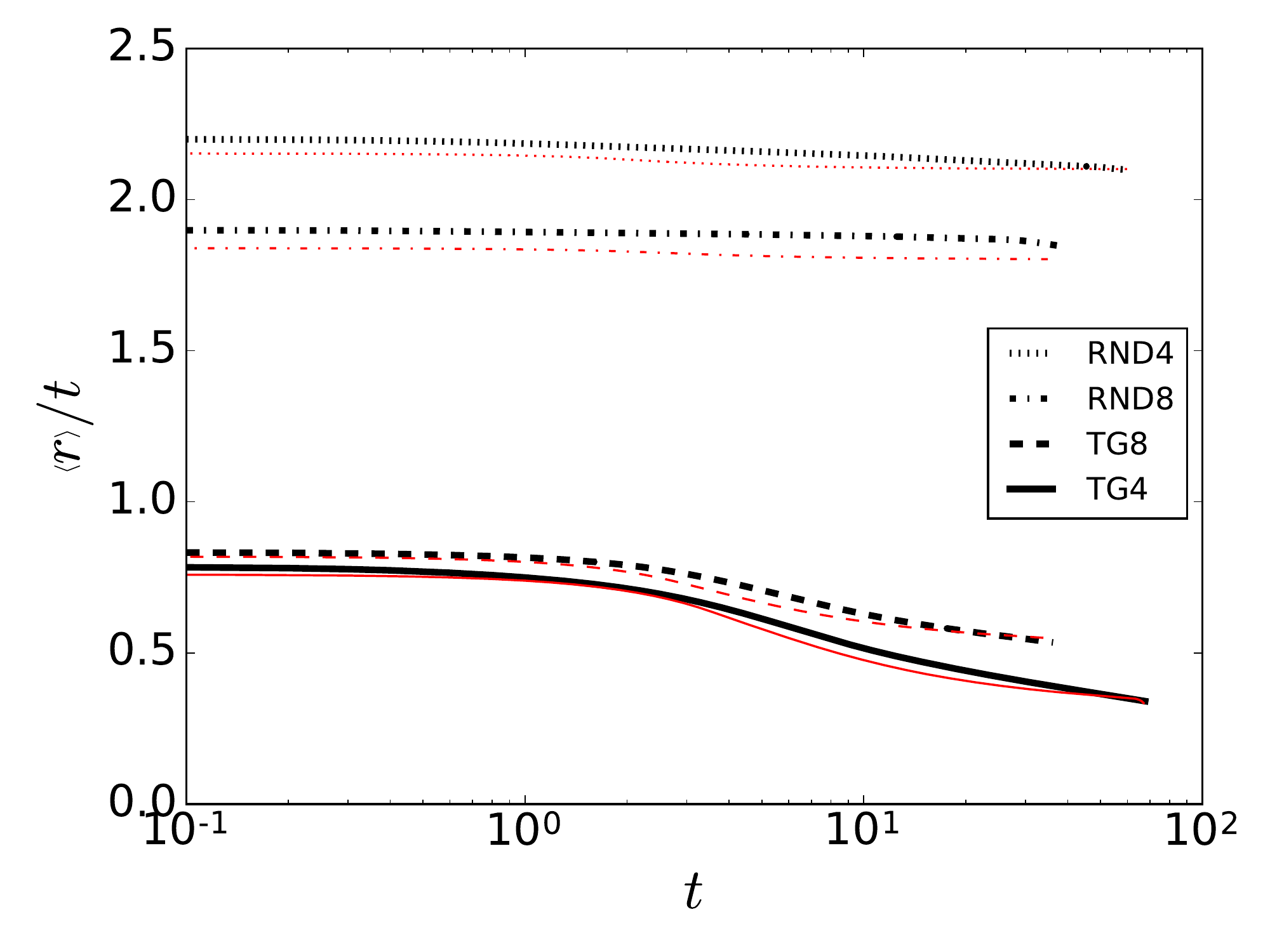}
\caption{{\it Left}: mean horizontal quadratic dispersion for
    all runs (thick black curves), normalized by the product of the
    mean squared horizontal velocity $u_{\perp}^2$ and the squared
    Lagrangian turn-over time $T_{l}^2$, as a function of time
    normalized by $T_{l}$. The same quantity obtained from the model
    is shown by the thin (red) curves, using the same line labels as
    their corresponding simulation. The mean quadratic dispersion
    obtained from the model is in good agreement with all
    simulations. {\it Right}: mean horizontal displacements normalized
    by the time for all runs (thick black curves), such that curves
    are flat when $\left< r \right> \sim t$. Again, 
    $\left< r \right>/t$ obtained from the model is shown in thin
    (red) curves.}
\label{fig:rt}
\end{figure}

Figure \ref{fig:prt} shows the probability density function $P(r,t)$
of a particle moving a horizontal distance $r$ after a time interval
$t$, computed using all simulations and all available time increments. 
The insets in Fig.~\ref{fig:prt} show the mean horizontal displacement
$\left<r \right>$ as a function of time (i.e., the first order moment
of the PDFs), while Fig.~\ref{fig:rt} shows the mean horizontal
quadratic displacement for all particles as a function of time, 
$\left< r ^2 \right>/(u_{\perp}^{2}T_{l}^{2})$ (i.e., the second
order moment of the PDFs in Fig.~\ref{fig:prt}), normalized by the
mean squared perpendicular velocity and the Lagrangian turnover
time using the scaling proposed in
\cite{aartrijk_single-particle_2008}. The PDFs and the displacements 
are different depending on the forcing and on the time scale
considered. At early times all simulations display  
$\left< r \right> \sim t$ and 
$\left< r^2 \right>/(u_{\perp}^{2}T_{l}^{2}) \sim (t/T_l)^2$, with
all curves in Fig.~\ref{fig:rt} collapsing in agreement with the
scaling observed in \cite{aartrijk_single-particle_2008}. This
indicates ballistic behavior at early times as in HIT
\cite{rast_turbulent_2016}. However, at late times ($t/T_{l}>1$) the
behavior of the TG and RND runs is clearly different, and differs
from the scaling proposed in
\cite{aartrijk_single-particle_2008}. In the TG runs, 
$\left< r \right>$ slows down but increases faster than 
$\sim t^{1/2}$ (see the insets in Fig.~\ref{fig:prt} and also the
mean horizontal displacements normalized by the time 
$\left< r \right>/t$ in Fig.~\ref{fig:rt}). This behavior at late
times is not universal, as the dependence of $\left< r \right>$ with
$t$ clearly varies with the level of stratification and with the
forcing. The case of RND forcing (see Figs.~\ref{fig:prt} and
\ref{fig:rt}) is even more interesting: $\left< r \right> \sim t$ and
$\left< r^2 \right>/(u_{\perp}^{2}T_{l}^{2}) \sim (t/T_l)^2$ even at
late times, and the maximum of $P(r,t)$ in $r$ (see the PDFs in
Fig. \ref{fig:prt}) displays a linear drift as $t$ increases.

\begin{figure}
\centering
\includegraphics[width=8.2cm]{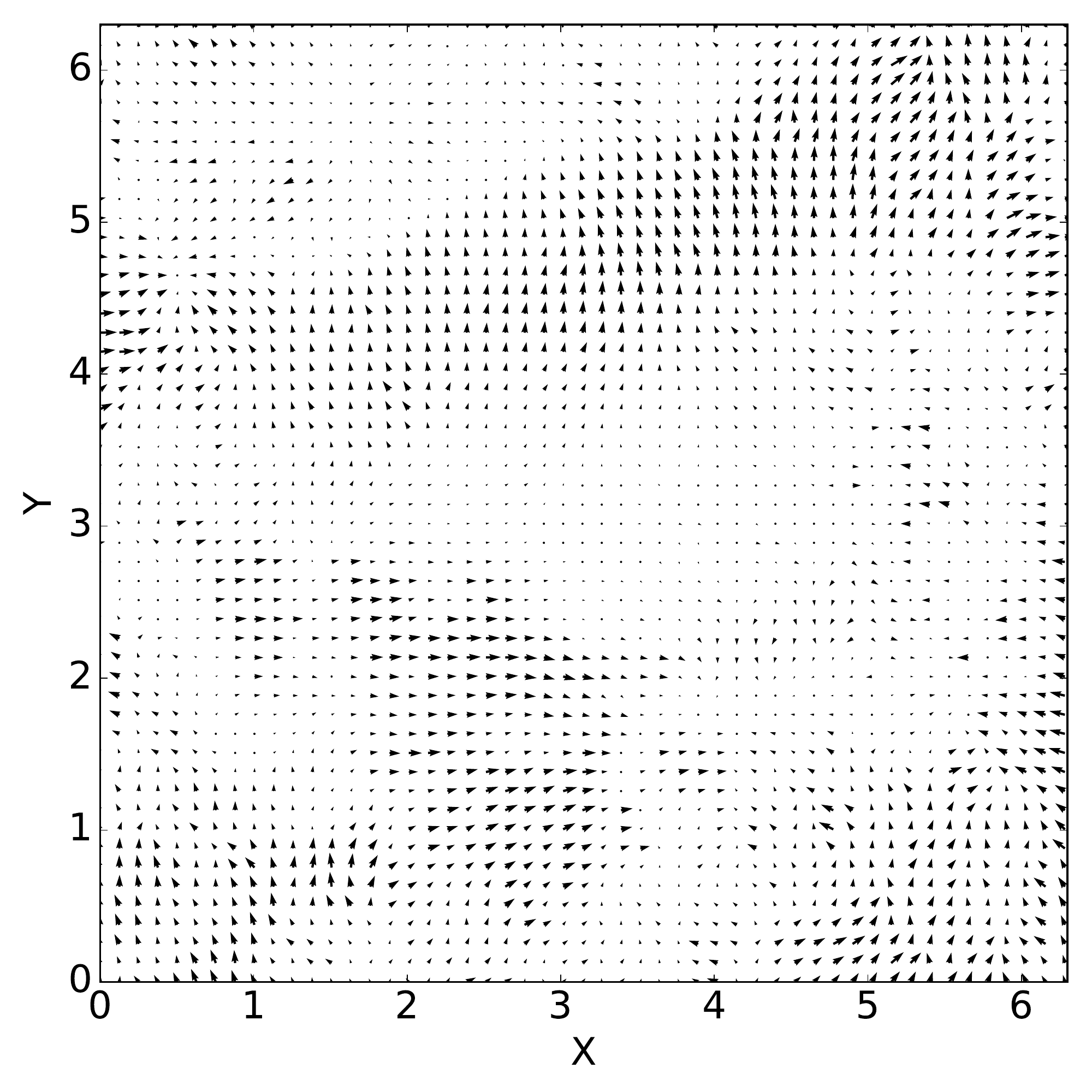}
\includegraphics[width=8.2cm]{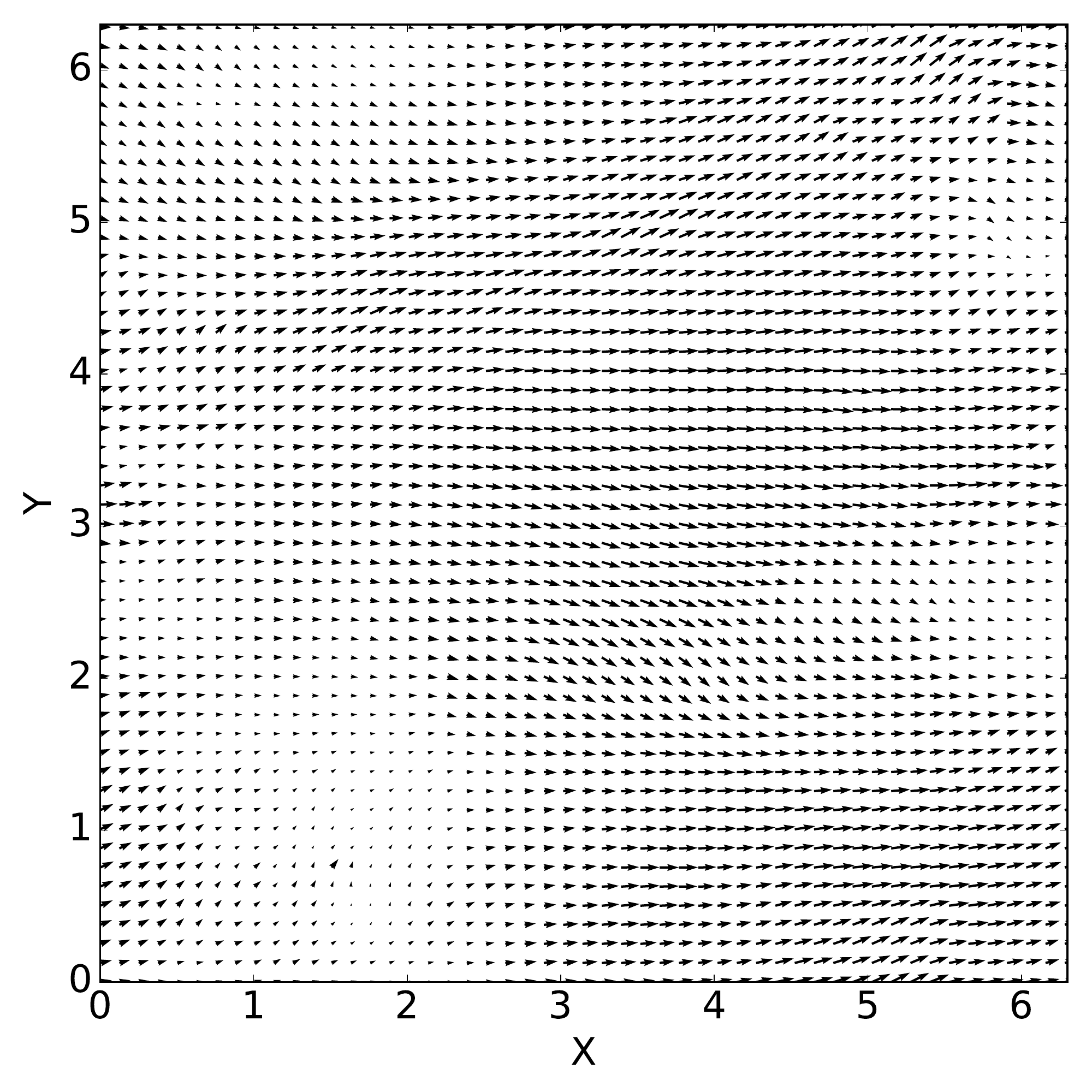}
\caption{\ADDB{Horizontal cuts (in the $x$-$y$ plane, and at $z=0$) of
    the horizontal velocity at time $t=20$ for ({\it left:)} the TG4
    run, and ({\it right:}) the RND4 run. Note the clear mean wind in
    the latter case, pointing approximately in the $\hat{x}$
    direction. Other layers (i.e., for other values of $z$) have mean
    horizontal winds pointing in other directions.}}
\label{fig:vperp_cuts}
\end{figure}

\begin{figure}
\centering
\includegraphics[width=8.5cm]{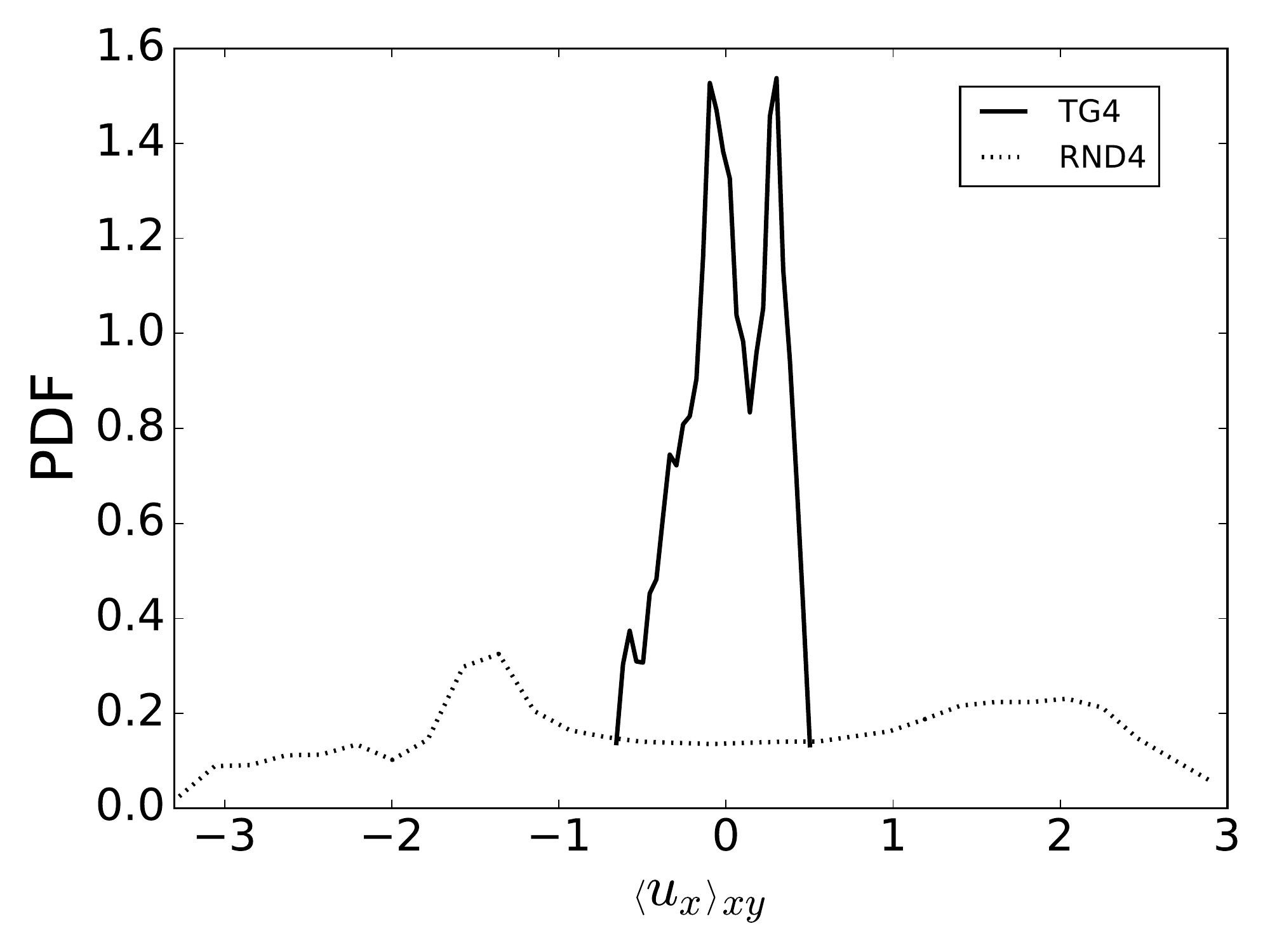}
\includegraphics[width=8.5cm]{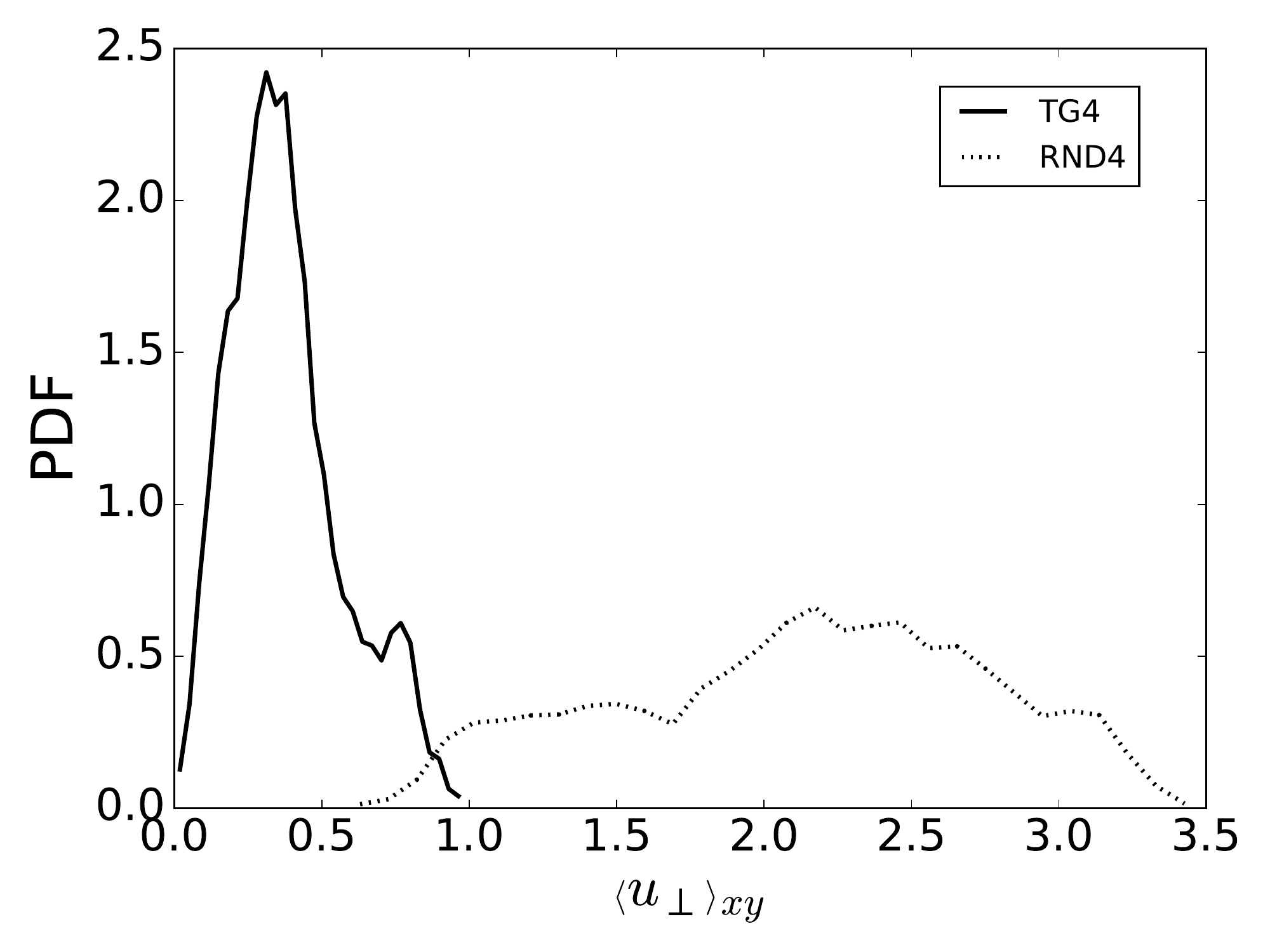}
\caption{\ADDB{PDFs of the mean horizontal winds for the TG4 and RND4
    runs. ({\it Left:}) PDF of $\left<u_x\right>_{xy}$, i.e., of the
      velocity in the $\hat{x}$ direction, $u_x$, averaged over the
      $x$ and $y$ coordinates, or equivalently, over planes at
      constant height. ({\it Right:}) Same for the r.m.s.~horizontal
      velocity $u_{\perp} = (u_{x}^{2}+u_{y}^{2})^{1/2}$, also
      averaged over planes at constant height.}}
\label{fig:VSHW}
\end{figure}

In the horizontal case, the main difference between TG and RND runs is
in the strength of the VSHW. In stratified flows, the anisotropic
energy transfer towards modes with $k_\perp \approx 0$ results in the 
formation of strong horizontal winds with vertical shear
\cite{smith_generation_2002, marino_large-scale_2014}. The flow is
then given by weakly coupled horizontal layers, each with a mean
horizontal velocity pointing in some direction. For RND forcing these
winds develop in the entire domain, resulting in the almost ballistic
motion of the particles observed in Fig.~\ref{fig:xyz} (each particle
is in a different layer, and thus the mean drift points in a different
direction). However, in the TG case the coherent forcing imposes a
large-scale structure that prevents the formation of mean winds,
except in the few horizontal layers where shear is maximum and the
forcing approaches zero. \ADDB{To illustrate this,
  Fig.~\ref{fig:vperp_cuts} shows horizontal cuts of the horizontal
  velocity for runs TG4 and RND4 (for the sake of clarity, the
  velocity at $1\%$ of the grid points in the horizontal plane is
  shown). For RND4, on the average the horizontal velocity points
  towards a well defined direction, generating a coherent drift for
  all particles spending a sufficiently long time in this plane. For
  this run (as well as for the RND8 run), the same behavior is
  observed in other horizontal cuts at different heights. However, for
  the TG4 run (as well as for TG8, not shown) no clear mean wind is
  observed. As mentioned above, small mean winds can only develop for
  TG forcing in a few horizontal layers. This is further confirmed in
  Fig.~\ref{fig:VSHW}, which shows the PDFs of the $\hat{x}$ component
  of the velocity $u_x$ (similar results are obtained for $u_y$), and
  of the r.m.s.~horizontal velocity $u_{\perp}$, averaged over horizontal
  planes before computing the PDFs (the PDFs are also averaged in
  time). These PDFs thus quantify the probability of finding
  horizontal layers with mean horizontal winds, in particular, for
  runs TG4 and RND4. Note the RND4 run has stronger mean horizontal
  winds when compared with the TG4 run, while the TG4 run has larger
  probability of having layers with zero mean velocity, further
  confirming the observations made in Fig.~\ref{fig:vperp_cuts}.}

Thus, we can conclude that in Fig.~\ref{fig:rt} there is a competition
between transport and trapping by turbulent eddies (which is
responsible for the mixing in the inertial range of HIT, and is also
visible in Fig.~\ref{fig:xyz}) and the coherent drift (due to the VSHW
in stratified turbulence). In the RND set of simulations the drift
dominates the motion of all particles giving ballistic-like behaviour
for all times, while in the TG set, as the drift is smaller and
affects only a fraction of the particles, the competition results in a
scaling in between ballistic and that observed in HIT. The VSHWs (and
their different strengths depending on the level of stratification and
type of forcing) can thus be expected to play an important role in the
diffusion of particles, as observed in the atmosphere
\cite{saffman_effect_1962,smith_role_1965}. The model we present next
confirms this.

To capture the observed horizontal single-particle dispersion we use
a continuous-time random walk model \cite{rast_turbulent_2016}, and
extend it to build a model that can capture both trapping of particles
by eddies as well as the drift caused by the VSHW. The motivation to
use a CTEC random walk model is the following. While in HIT dispersion
of particles is ballistic for short times and diffusive 
($\left< r \right> \sim t^{1/2}$) for late times, it has been observed
that the PDFs $P(r,t)$ show deviations from a Rayleigh
distribution. In particular, at early times $P(r,t)$ displays a
slightly smaller probability of particles having large displacements
from their origins when compared with a Rayleigh
distribution, while at late times a small excess of large
displacements can be observed. \ADDB{This indicates that a simple
  random walk model is insufficient to capture the dynamics of the
  system.} Using a point vortex model \cite{rast_2011}, the early time
behavior was shown to be associated with trapping of particles by
the eddies, which then cannot travel as far as they could in a
random walk. Naturally, the trapping time is a continuous random 
variable, whose distribution can be obtained, e.g., from comparisons
with a point vortex model \cite{rast_turbulent_2016}. It is clear
from Fig.~\ref{fig:xyz} that trapping by eddies also plays a role in
the horizontal displacement in our simulations, at least in the runs
in which VSHW are not dominant. Thus, in our random walk model each
particle is trapped in an eddy and displaces 
\begin{equation}
dr_{t}=2 r_{t} |\sin(\theta_{t})| 
\end{equation}
for a time $t_{t}$, where $r_{t}$ is the radius of the eddy, and 
\begin{equation}
\theta_{t} = U_{t} t_{r} /r_{t} 
\end{equation}
is the central angle of motion of the particle while trapped, where
$U_{t}$ is the Lagrangian particle velocity.  The particle enters each
eddy (i.e., each trap) with random phase. The model has no free
parameters, and the probability distributions of $U_{t}$, $t_{t}$, and
$r_{t}$ (with the sequence of random values for these quantities
corresponding to the successive motion of a particle from one eddy to
the next) are obtained from observations or from Kolmogorov theory of
turbulence as follows. The probability density of finding an eddy of
radius $r_{t}$ is taken to be Kolmogorovly distributed 
\begin{equation}
P(r_{t}) \sim r_{t}^{4/3} 
\end{equation}
for $r_{t}<L/2$, where $L$ is the Eulerian integral length of the
flow as defined above. \ADDB{The assumption that the eddy distribution
  is Kolmogorovian is based on observations that the perpendicular
  velocity in stably stratified turbulence follows Kolmogorov
  scaling with the perpendicular wavenumber \cite{riley_dynamics_2003,
    waite_stratified_2004, lindborg_energy_2006,
    brethouwer_scaling_2007}.} The probability density of a given
trapping time for a particular step $t_{t}$ is uniform between $0$ and
$T_{e}$, where $T_{e}$ is the Eulerian turnover time
\cite{rast_2011}. Finally, the probability density of particle
velocities $P(U_{t})$ is obtained from the PDF of the Lagrangian 
perpendicular velocity in the simulation (after subtracting the mean 
velocity associated with the drift caused by the VSHW). In each step
of the random walk, a set of these variables is randomly generated,
each chosen independently, and their values are kept constant over the
trap duration $t_{t}$.

To this eddy-constrained random walk, a uniform drift $D_t = W t_t$ is
added to each particle, with the PDF of the wind $W$ (different for
each particle) given by a bimodal Gaussian distribution corresponding
to the best fit to the PDF of the VSHW \ADDB{(see
  Fig. \ref{fig:VSHW})} in each simulation. Note the VSHWs are
different depending on the forcing, with larger values of $W$ in
the RND runs, and lower values (and larger probability of having
particles with $W\approx 0$) in the TG runs. The random walk is then 
constructed as the sum over steps of length $dr_t + D_t$.

\begin{figure}
\centering
\includegraphics[width=8.9cm]{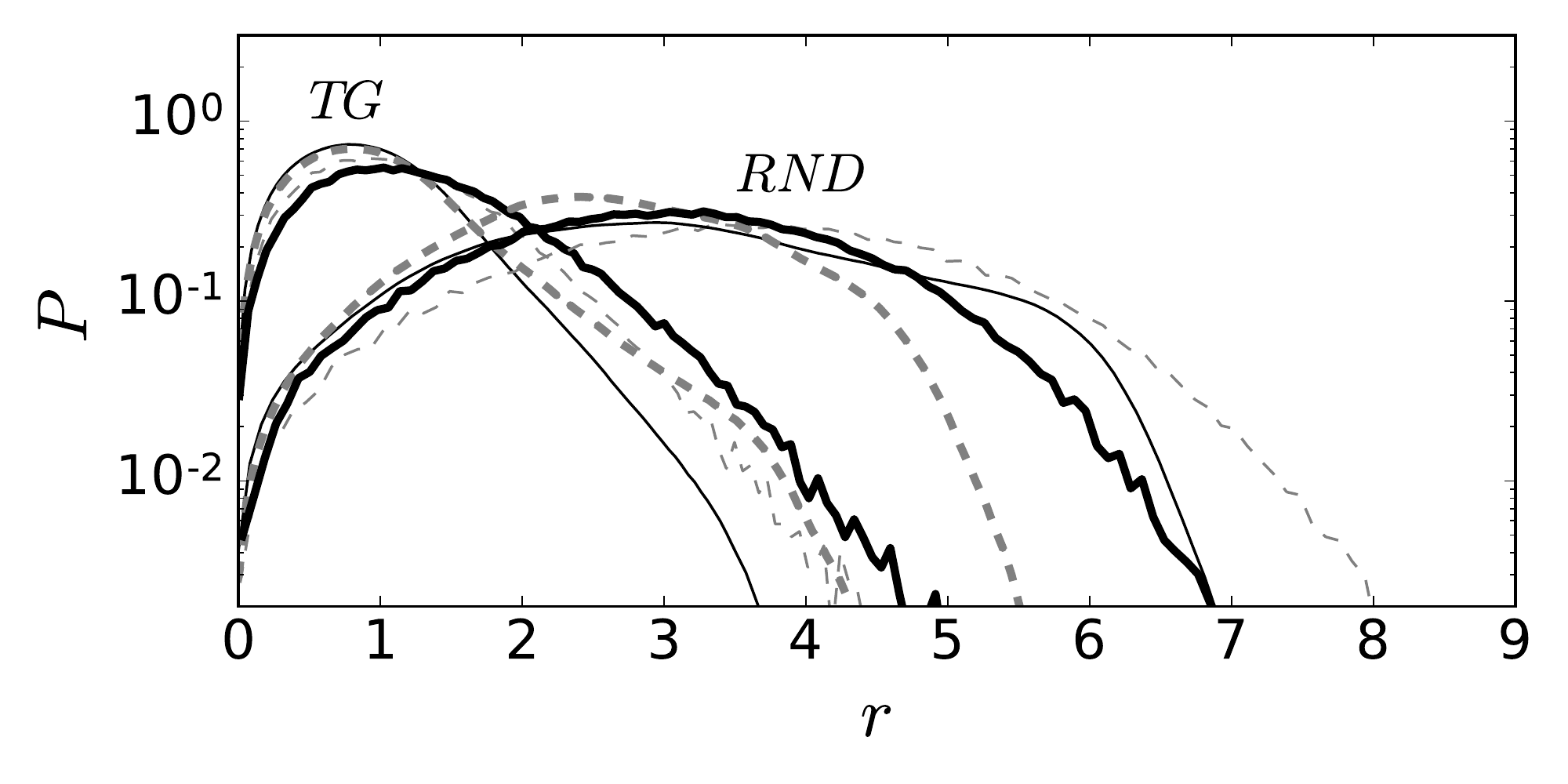}
\includegraphics[width=8.9cm]{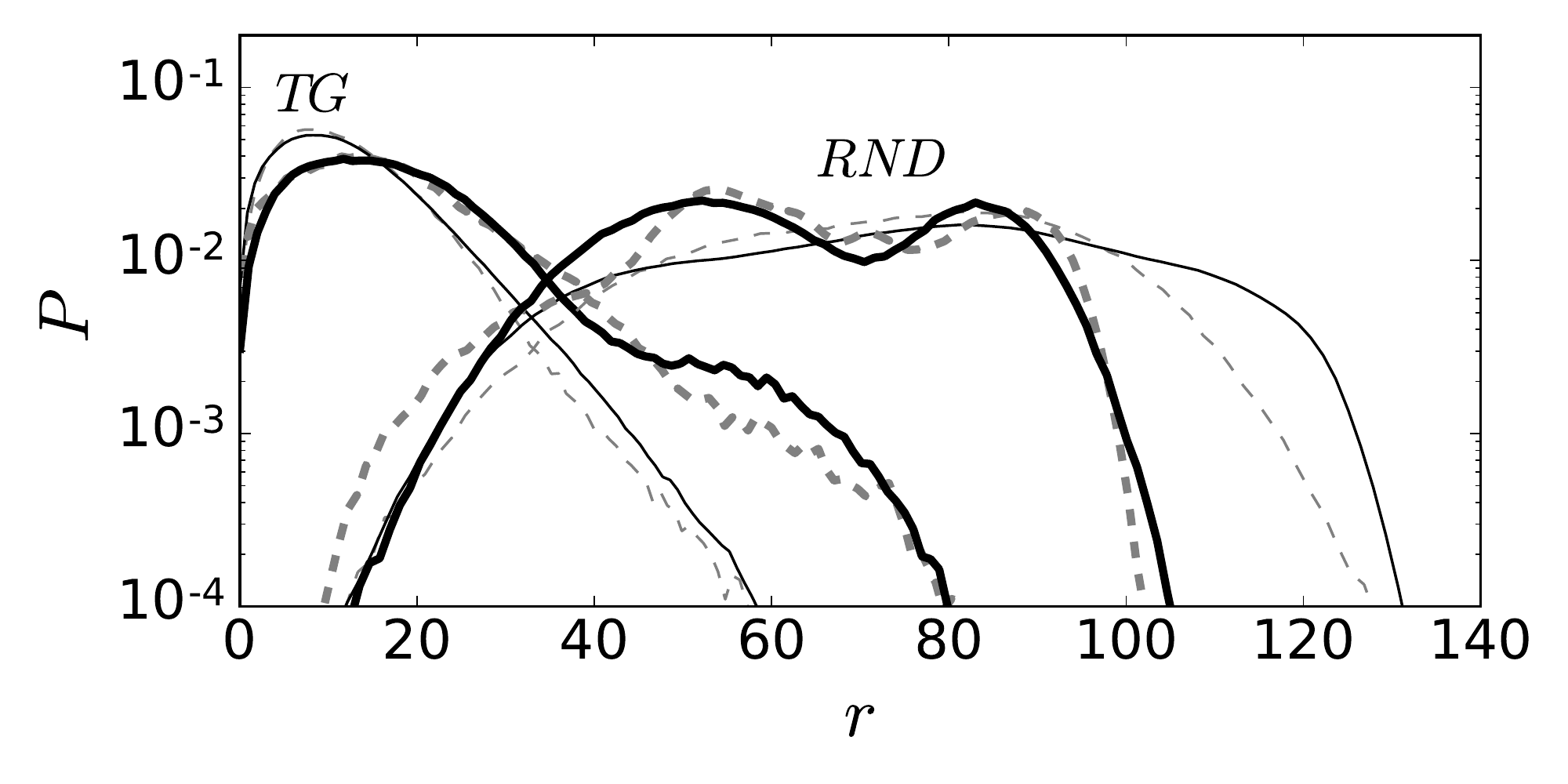}
\caption{$P(r,t)$ at different times for the simulations (solid lines)
  and for the model (dashed lines). Thick lines correspond to flows
  with $N=4$, and thin lines to $N=8$. PDFs peaked at small values of
  $r$ are for TG forcing, while PDFs peaked at large values of $r$ are
  for RND forcing (as labeled). {\it Left}: $t=1.5$ ($t<T_{e}$). 
  {\it Right}: $t=35$ ($t>T_{l}$).}
\label{fig:pr_tfijo}
\end{figure}

The modeled $P(r,t)$ are also shown in Fig.~\ref{fig:prt}, which are
in good agreement with simulations. Note also that the model
captures differences between TG and RND runs, differences between
simulations with different values of the Brunt-V\"{a}is\"{a}l\"{a}
frequency $N$, as well as the anomalous behavior of $\left<r\right>$
and of $\left<r^2\right>$  (Fig. \ref{fig:rt}) at late times in all
cases. The PDFs from the model and the simulations at two fixed times 
are also compared in Fig.~\ref{fig:pr_tfijo}. It is clear the PDFs are
not Rayleigh, confirming turbulent transport in stratified flows
cannot be modeled simply as a standard random walk. Moreover, two
regimes are found for times shorter than the Eulerian turnover
time $T_e$ and for times larger than the Lagrangian turnover time
$T_l$. For $t<T_e$ trapping is more relevant than the drift, while
for $t>T_l$ the drift dominates the dispersion giving very different
PDFs in the TG and RND cases. Note that unlike HIT, the
Lagrangian turnover times in all these simulations are larger than the
corresponding Eulerian times, a result of a long term correlation in
each particle trajectory caused by the drift by the VSHW. Indeed, the
RND runs (which have stronger VSHW) have a larger separation between
$T_l$ and $T_e$, and their separation also increases with increasing
Brunt-V\"{a}is\"{a}l\"{a} frequency $N$.

\section{Conclusions \label{sec:conclusions}}

In this paper we studied single-particle dispersion in stably
stratified turbulence using different forcing functions and
Brunt-V\"{a}is\"{a}l\"{a} frequencies. We showed that vertical
dispersion is strongly reduced by the stratification, with the
vertical Lagrangian velocity following a spectrum compatible with
observations from wave-dominated geophysical flows
\cite{garrett_space-time_1975, lien_lagrangian_1998,
  dasaro_lagrangian_2000}. Knowledge of this spectrum is enough to
construct a random superposition of internal gravity waves which
gives probability distribution functions of the waiting times of the
Lagrangian particles in good agreement with the data, and mean
vertical displacements in good agreement with the simulations up to
the Brunt-V\"{a}is\"{a}l\"{a} period in all simulations and for
longer times in the simulations with random forcing. We also showed
that horizontal dispersion differs from HIT and is strongly
influenced by the large scale vertically sheared horizontal winds
generated by the stratification \cite{smith_generation_2002,
  marino_large-scale_2014, clark_di_leoni_absorption_2015}. 
Knowledge of the Eulerian typical flow velocity, of the Eulerian
integral length, of the probability density function of the horizontal
Lagrangian velocity, and of the strength of the horizontal winds is
enough to build a continuous-time eddy-constrained random walk model,
which takes into account particle trapping by eddies and drift by the
mean winds, and which correctly reproduces the probability density
function of the horizontal particle displacements in all simulations.

\ADDA{This model assumes that vertical displacements of the Lagrangian
  particles are negligible, or at least that they remain small across
  the time scales associated with the horizontal displacements, in
  such a way that each particle can be treated as being transported
  along a unique horizontal layer with a given horizontal wind. This
  is in good agreement with observations in our simulations, as well
  as with previous studies of Lagrangian transport and dispersion in
  stably stratified flows \cite{nicolleau_turbulent_2000,
    aartrijk_single-particle_2008}. In particular, the observed mean
  squared vertical displacements confirm there is little mixing after
  one Brunt-V\"{a}is\"{a}l\"{a} wave period, with saturation in the
  case of isotropic random forcing, and with a slow increase in the
  case of Taylor-Green forcing. However, it may be the case that
  for stronger turbulence, or for weaker stratification (i.e., for
  larger values of the buoyancy Reynolds number), vertical mixing can
  be stronger and cause particles to change layers on faster time
  scales. Moreover, our results for two different forcing functions
  indicate that the local gradient Richardson number may be a better
  indicator of such a change in behavior. A detailed parametric study
  of vertical transport for moderate values of the Froude number and
  for larger Reynolds numbers is left for future work.}

For the range of parameters considered in this study, the results show
that single-particle dispersion in stably stratified fluids departs
significantly from the behavior observed in HIT, which has 
strong implications for the study of transport in geophysical
flows. In particular, transport at times larger than the Lagrangian 
turnover time is not universal, and strongly dependent on the presence
or not of horizontal winds. Thus, even the mean displacement of the
particles cannot be modeled by a single power law of time, and may not
be self-similar. As mentioned above, the models we presented capture
this behavior and yield results in good agreement with numerical
simulations. For the range of parameters considered here, the model
indicates that while vertical transport is mostly mediated by a random
superposition of internal gravity waves, proper capturing of the
horizontal transport requires a superposition of a random walk with
trapping, with a mean drift caused by horizontal winds, which are
dependent on the forcing, on the stratification level, and in
geophysical scenarios can also be affected by topography and other
factors. Finally, the model provides a statistical prediction of
moments of the PDF of dispersion without the need of explicit
simulation of the turbulent flow and of individual particle
trajectories. Note this is of particular importance for the modeling
of geophysical flows. While in recent years state of the art
simulations allowed studies with very high spatial resolution of
atmospheric and oceanic flows, forecasts and large scale modeling
require ensembles of runs which are performed at lower resolutions and
which cannot resolve the small scale turbulence. As a result, the
development of statistical models not only provides a deeper insight
into the mechanisms behind particle dispersion, but can also allow the
development of subgrid scale models of turbulent particle transport
and mixing.

\begin{acknowledgments}
N.E.S.~and P.D.M.~acknowledge support from UBACYT Grant
No.~20020130100738BA, and PICT  Grants Nos.~2011-1529 and 
2015-3530. P.D.M. also acknowledges support from the CISL visitor
program and from the Geophysical Turbulence Program at NCAR.
\end{acknowledgments}

\appendix
\section{Horizontal dispersion with Kolmogorov forcing}

\begin{figure}
\centering
\includegraphics[width=8.0cm]{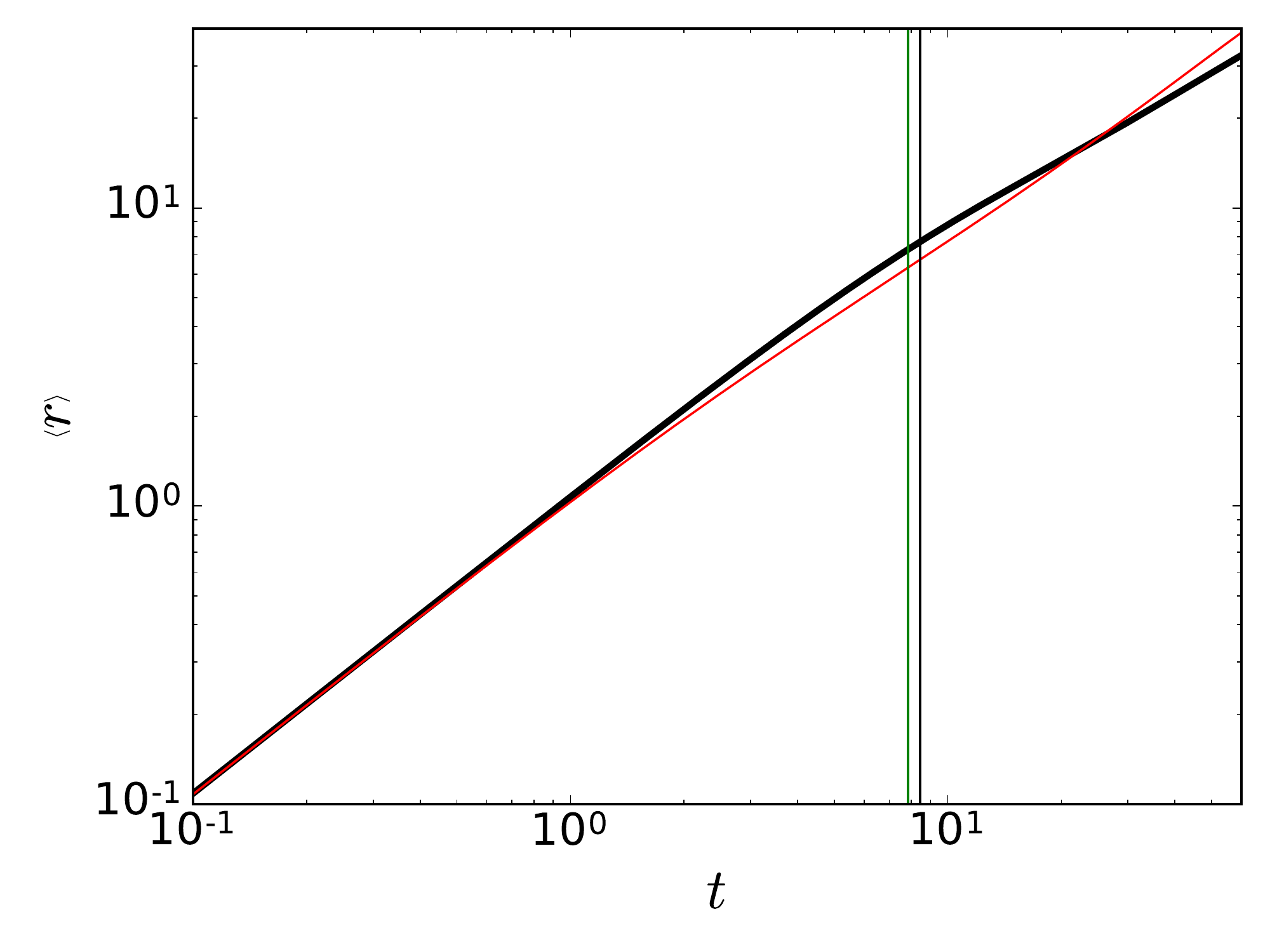}
\includegraphics[width=8.0cm]{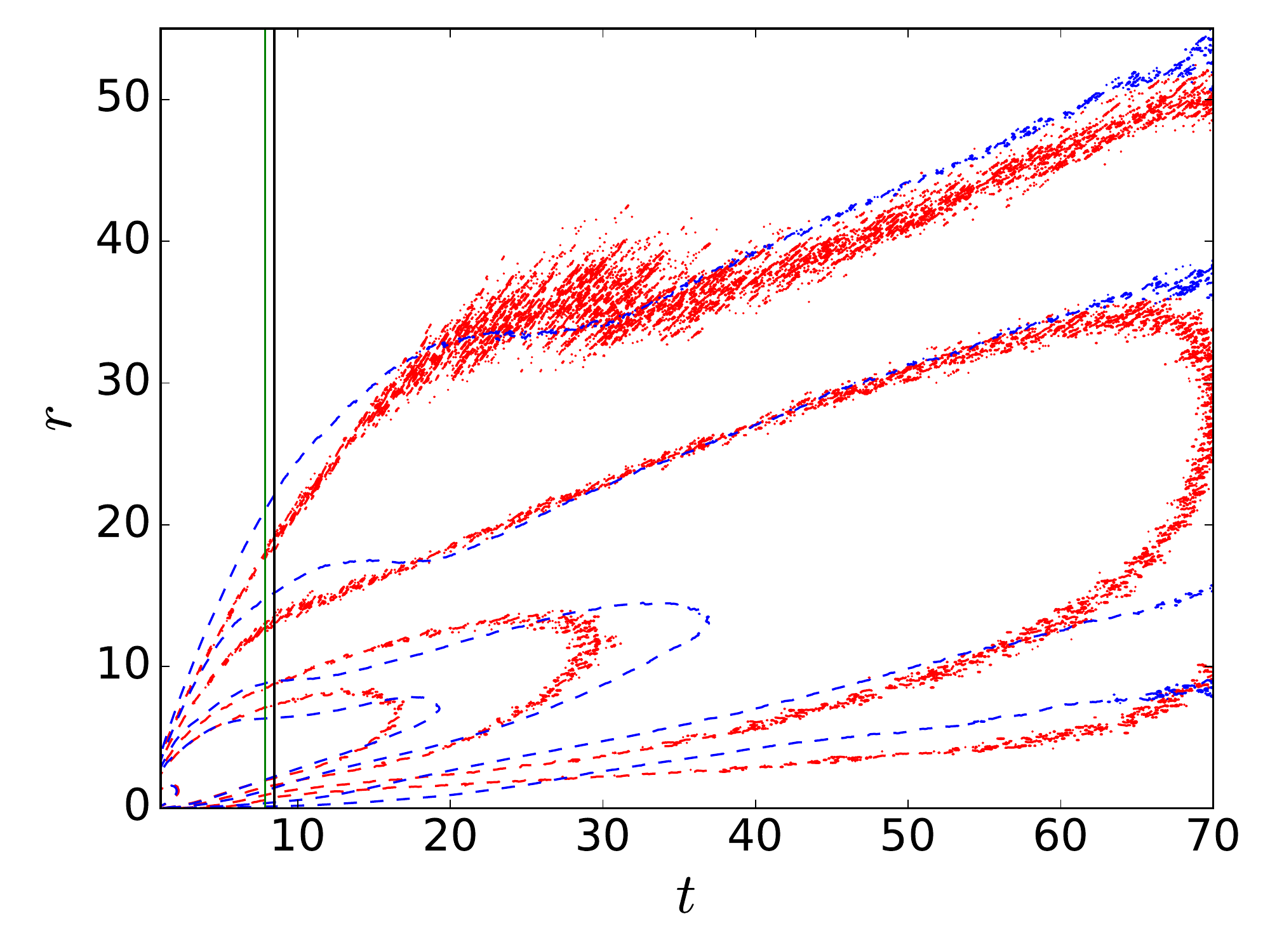}
\caption{{\it (Color online)} {\it Left}: Mean horizontal
    dispersion as a function of time for a run with Kolmogorov forcing
    and $N=8$ (thick black line). Dispersion in the random walk
    model is shown by the thin (red) line. {\it Right}: Isocontours of
    $P(r,t)$ for the simulation (dashed blue curves) and for the model
    (dotted red curves). Vertical lines (from left to right) indicate
    respectively the Eulerian $T_{e}$ and Lagrangian $T_{l}$ turnover
    times.}
\label{fig:kol}
\end{figure}

To further test the model for horizontal dispersion we briefly
present here a $512^3$ simulation with Brunt-V\"{a}is\"{a}l\"{a}
frequency $N=8$, the same parameters as in the other simulations 
($\nu=\kappa=8 \times 10^{-4}$ and Schmidt number 
$\textrm{Sc} = 1$), but with Kolmogorov forcing. To force at the
same scales as in the previous simulations, we apply the Kolmogorov
forcing at $k=1$ and $2$ (see, e.g., \cite{rollin2011}) and use
\begin{equation}
  {\bf f} = f_{0} \left[ \sin(y) + \sin(2y) \right] \hat{x} .
\label{eq:kolmogorov_forcing}
\end{equation}
While in RND forcing all three components of the velocity field are
forced and the forcing is isotropic, and in TG forcing both
horizontal components of the velocity are forced, in Kolmogorov
forcing only one component of the velocity is forced, resulting in
the turbulent steady state in an anisotropic flow even in the
horizontal plane. As an example, in the turbulent steady state the
r.m.s.~Eulerian velocity in the $x$ direction is almost three times
larger than the r.m.s.~value of the Eulerian velocity in the $y$
direction.

For this simulation, as for the previous simulations, we studied
the mean horizontal dispersion and the probability distribution
$P(r,t)$ of finding a particle at a given distance $r$ at time $t$
from its original location (see Fig.~\ref{fig:kol}). Using the
typical value of the horizontal Eulerian velocity (averaged in the
$x$ and $y$ directions), the Eulerian turnover time, the PDF of the
Lagrangian velocity of the particles, and the typical amplitude of
the horizontal winds, we also computed the random walk model (see
also Fig.~\ref{fig:kol}). Considering the strong horizontal
anisotropy in this flow, which is different from the other forcing
functions studied in this work, the model is in good agreement with 
the simulations, especially for the PDF of the displacements
$P(r,t)$.

The mean horizontal dispersion in the simulation shows ballistic
behaviour at early times ($\left< r \right> \sim t$), and slows down
at later times, albeit less than in the TG simulations. Again, there
is a competition between diffusion and trapping by the turbulent
eddies and the drift by the horizontal winds, which in this simulation
are weaker than in the RND runs but stronger than in the TG
runs. The PDF $P(r,t)$ has an excess of particles that travel far
from their origin at early times (up to $t \approx 30$), which can
be seen as a bump in the isocontours. Then, horizontal winds give
almost linear isocontours in time. The model, without any
modification, can capture for this flow both features and the
overall shape of the PDF.

\bibliography{ms}

\end{document}